\documentclass[lettersize,journal]{IEEEtran}
\usepackage{amsmath,amsfonts}
\usepackage{algorithmic}
\usepackage{algorithm}
\usepackage{array}
\usepackage[caption=false,font=normalsize,labelfont=sf,textfont=sf]{subfig}
\usepackage{textcomp}
\usepackage{stfloats}
\usepackage{url}
\usepackage{verbatim}
\usepackage{cite}
\pdfpageattr{/Group << /S /Transparency /I true /CS /DeviceRGB >>}

\usepackage[utf8]{inputenc}
\usepackage[T1]{fontenc}
\usepackage{textcomp}
\usepackage{microtype}
\usepackage{calc}
\usepackage[normalem]{ulem}

\usepackage{lipsum}

\usepackage[range-phrase=--,per-mode=symbol-or-fraction,binary-units=true,range-units=single,list-units=single,detect-all,forbid-literal-units=true]{siunitx}
\AtBeginDocument{
\DeclareSIUnit\bit{bit}
\DeclareSIUnit\byte{Byte}
\DeclareSIUnit\decibelm{dBm}
\DeclareSIUnit\vehicle{veh}
}

\usepackage{silence}
\usepackage{csquotes}

\usepackage{amsmath}
\usepackage{amssymb}
\usepackage{amsfonts}
\usepackage{amsthm}

\usepackage{mathtools}

\usepackage{booktabs}
\usepackage{url}

\usepackage[inline]{enumitem}

\usepackage[pdftex]{graphicx}
\DeclareGraphicsExtensions{.pdf,.png,.jpg}
\usepackage{tikz}
\usetikzlibrary{arrows}
\usetikzlibrary{calc}
\usetikzlibrary{chains}
\usetikzlibrary{scopes}

\usepackage[american]{babel}
\usepackage{hyphenat}

\usepackage{algorithmic}
\usepackage{algorithm}

\usepackage[capitalize,noabbrev,nameinlink]{cleveref}

\RequirePackage{xstring}
\RequirePackage{xparse}
\RequirePackage[]{acro}
\makeatletter
\@ifpackagelater{acro}{2019/02/17}{
	\NewDocumentCommand\acrodef{mO{#1}mG{}}{\DeclareAcronym{#1}{short={#2}, long={#3}, foreign-plural={}, #4}}
}{
	\NewDocumentCommand\acrodef{mO{#1}mG{}}{\DeclareAcronym{#1}{short={#2}, long={#3}, #4}}
}
\makeatother
\acrodef{AD}{Access Delay}
\acrodef{AIFS}{Arbitration Inter-Frame Spacing}
\acrodef{AoI}{Age of Information}
\acrodef{AWGN}{Additive White Gaussian Noise}
\acrodef{BS}{Base Station}
\acrodef{BSS}{Basic Service Set}
\acrodef{CAM}{Cooperative Awareness Message}
\acrodef{CPM}{Collective Perception Message}
\acrodef{CBF}{Contention-Based Forwarding}
\acrodef{CBR}{Channel Busy Ratio}
\acrodef{CDF}{Cumulative Distribution Function}
\acrodef{CCDF}{Complementary Cumulative Distribution Function}
\acrodef{CDMA}{Code Division Multiple Access}
\acrodef{CCH}{Control Channel}
\acrodef{CSMA}{Carrier-Sense Multiple Access}
\acrodef{C-ITS}{Cooperative Intelligent Transportation System}{short-plural-form={C-ITS}}
\acrodef{DENM}{Decentralized Environmental Notification Message}
\acrodef{DIFS}{DCF Inter-Frame Space}
\acrodef{DSRC}{Dedicated Short-Range Communication}
\acrodef{DTMC}{Discrete Time Markov Chain}
\acrodef{FCFS}{First Come First Served}
\acrodef{FDMA}{Frequency Division Multiple Access}
\acrodef{GFMA}{Grant-Free Multiple Access}
\acrodef{IoT}{Internet of Things}
\acrodef{ITS}{Intelligent Transportation System}{long-plural-form={}}
\acrodef{IVC}{Inter Vehicle Communication}
\acrodef{LCFS}{Last Come First Served}
\acrodef{LCFSwO}{Last Come First Served with Overwrite}
\acrodef{LDM}{Local Dynamic Map}
\acrodef{M2M}{Machine-to-Machine}
\acrodef{MAC}{Medium Access Control}
\acrodef{MCM}{Maneuver Coordination Message}
\acrodef{MIMO}{Multiple Input Multiple Output}
\acrodef{MPR}{Multi-Packet Reception}
\acrodef{MTC}{Machine Type Communications}
\acrodef{NOMA}{Non-Orthogonal Multiple Access}
\acrodef{OBU}{On-Board Unit}
\acrodef{OFDMA}{Orthogonal Frequency Division Multiple Access}
\acrodef{OMA}{Orthogonal Multiple Access}
\acrodef{PDF}{Probability Density Function}
\acrodef{PDR}{Packet Delivery Ratio}
\acrodef{RSU}{Road Side Unit}
\acrodef{SA}{Slotted ALOHA}
\acrodef{SCH}{Service Channel}
\acrodef{SIC}{Successive Interference Cancellation}
\acrodef{SNR}{Signal to Noise Ratio}
\acrodef{SNIR}{Signal to Noise plus Interference Ratio}
\acrodef{V2I}{Vehicle-to-Infrastructure}
\acrodef{V2V}{Vehicle-to-Vehicle}
\acrodef{V2X}{Vehicle-to-Everything}
\acrodef{VANET}{Vehicular Ad Hoc Network}
\acrodef{VLC}{Visible Light Communication}
\acrodef{TDMA}{Time Division Multiple Access}
\acrodef{UAV}{Unmanned Aerial Vehicle}
\acrodef{URLLC}{Ultra-Reliable Low-Latency Communications}

\usepackage[color]{changebar}
\def\todoCtd#1{%
	TODO: #1%
	\ifx&#1&...\fi%
	\endgroup
	\cbend
	\relax
}

\NewDocumentCommand\IEEE{ s m >{\SplitArgument{4}{/}}d[] }{%
    \IfBooleanTF{#1}{}{IEEE\,}% suppress IEEE when using starred form
    \nolinebreak[2]% this is a somewhat bad place for a line break
    #2%
    \IfNoValueTF{#3}{%
    }{%
        \sommerIEEELettersSlashed#3%
    }%
}
\newcommand{\sommerIEEELettersSlashed}[5]{%
    \IfNoValueTF{#2}{%
    }{%
        \nolinebreak[3]% multiple letters, this is just a very bad place for a line break
    }%
	#1%
	\IfNoValueTF{#2}{}{/#2}%
	\IfNoValueTF{#3}{}{/#3}%
	\IfNoValueTF{#4}{}{/#4}%
	\IfNoValueTF{#5}{}{/#5}%
}

\newcommand{\matlab}{MATLAB\textsuperscript\textregistered}

\begin{document}

\title{Analysis of Status Update in Wireless Networks with Successive Interference Cancellation}

\author{Asmad Bin Abdul Razzaque,~\IEEEmembership{Graduate Student Member,~IEEE} and Andrea Baiocchi,~\IEEEmembership{Member,~IEEE}
\thanks{Asmad B.A. Razzaque is the corresponding author.}
\thanks{A.B.A Razzaque and A. Baiocchi are affiliated with the Dept.\ of Information Engineering, Electronics and Telecommunications (DIET), University of Rome Sapienza, Italy (asmadbin.razzaque@uniroma1.it, andrea.baiocchi@uniroma1.it).}}

%\author{Andrea Baiocchi,~\IEEEmembership{Member,~IEEE,}
        % <-this % stops a space
%\thanks{.}}

% The paper headers
\markboth{IEEE TRANSACTIONS ON WIRELESS COMMUNICATIONS, AUGUST, 2024}%
{Shell \MakeLowercase{\textit{et al.}}: Analysis of Status Update in Wireless Networks with Successive Interference Cancellation}

% Remember, if you use this you must call \IEEEpubidadjcol in the second
% column for its text to clear the IEEEpubid mark.

\maketitle

\begin{abstract}
Data collection in an IoT environment requires simple and effective communication solutions to address resource constraints, ensure network efficiency, while achieving scalability.
Efficiency is evaluated based on the timeliness of collected data (Age of Information), the energy spent per delivered unit of data, and the effectiveness in utilizing spectrum resources.
This paper addresses a random multiple access adaptive system, in which a large number of devices send sporadic messages in non-periodic pattern.
In particular, our analysis highlights the potential of Successive Interference Cancellation and identifies an adaptive parameter setting to maximize its benefits as the level of contention on the shared channel varies.
An analytical model is defined, easily scalable with the number of nodes and yielding all the relevant metrics.
Evidence of the accuracy of the model is given by comparing predicted results against simulations.
The model is utilized to assess the trade-off between Age of Information and energy consumption, revealing a sharp relationship between the two.
The considered approach lends itself to many generalizations and applications to massive machine-type communications and IoT networks.
\end{abstract}


\acresetall
\IEEEpeerreviewmaketitle

% -------------- Section end marker --------------
%                _       _
%               ( )_    ( )
%    ___  _   _ | ,_)   | |__     __   _ __   __
%  /'___)( ) ( )| |     |  _ `\ /'__`\( '__)/'__`\
% ( (___ | (_) || |_    | | | |(  ___/| |  (  ___/
% `\____)`\___/'`\__)   (_) (_)`\____)(_)  `\____)
%
% -------------- Section end marker --------------

\section{Introduction}
\label{sec:introduction}

\IEEEPARstart{I}{nternet} of things is a prominent application of next-generation wireless and cellular technologies.
The \ac{IoT} is mainly realized as a wireless sensor network with a large number of devices, from wearable to environmental sensors and industrial robots, designed to interact in various scenarios such as transportation systems, power grids, logistics, pollution monitoring, waste treatment, industrial production, agriculture, and food chains.
%The \ac{IoT} can be mainly realized as a wireless sensor network with a large number of sensor devices, from wearable devices to environmental sensors, industrial robots, actuators designed to interact in different scenarios, e.g., transportation systems, power grid, logistics, pollution monitoring, waste treatment, industry production lines, agriculture, and food chains.
Key common features of these applications, from a communication infrastructure point of view, are the usage of wireless medium and the need for easy sharing of communication channels by a large number of transmitting devices.
Another key feature of these applications is that traffic patterns are typically discontinuous, with random and sporadic data emissions to be delivered to a collection point such as a gateway or base station.
Typically, these messages are short, often just a few hundred bytes.
This scenario is also referred to as massive \ac{MTC}, especially in the context of 5G and 6G technologies \cite{Sharma2020,Huda2020_mMTC6G}.
As an example, the third generation partnership project (3GPP) has considered scenarios with a large number of \ac{MTC} devices to realize cellular-IoT technology for 6G wireless communication technologies \cite{Shahab2020grantfree}.
% \ac{MTC} devices require massive connectivity, which is a major challenge for next-generation wireless communication technologies.
However, multiple access methods used in cellular network face significant difficulties supporting massive \ac{IoT}.
These issues have been highlighted and several improvements have been proposed for LTE-A \cite{TelloOquendo2018}, 5G \cite{Pitaval2020,Zhang20235G}, and NB-IoT \cite{Chen2022NBIoT}.

Traditional multiple access techniques fall into two major categories.
A first class of multiple access techniques encompasses those specifically designed for reaping high spectral efficiency for a few simultaneously active users, each transmitting at a large bit rate (from Mbit/s to Gbit/s).
That is the case of, e.g., Wi-Fi.
A second class of multiple access techniques is designed for a large population of nodes, each transmitting infrequently, aiming at ease of channel sharing, rather than spectral efficiency.
The case of \ac{SA} is a paradigm of this approach, where the simplicity of implementation is traded off with a severe limitation in the maximum achievable stable throughput, at least for classic \ac{SA}.
To meet the requirements of \ac{MTC} devices, future wireless technologies call for new approaches to multiple access techniques.

Advances in hardware capabilities are facilitating the introduction of sophisticated coding schemes and signal processing algorithms that enable \ac{MPR} \cite{Lang2001}.
One major example of these advanced signal processing techniques is \ac{SIC} \cite{Miridakis2013,Sen2013SIC}.

\ac{SIC} enables the decoding of multiple superposed messages, by removing the interference caused by those messages that have been successfully decoded first.
Several \ac{SIC} algorithms have been proposed \cite{Sen2013SIC,Zhang2014SIC,Chen2019SIC,Abu2019, Narayanasamy2021}.
More generally, in the realm of multiple access techniques, \ac{MPR} has emerged as a central focus for managing the interference.
It tackles the complex challenges of spectrum usage and interference reduction, aiming to enhance connectivity, network performance, and user experience as we progress towards the 6G era of machine-type communications \cite{Siddiqui2023,Miridakis2013,mahmood2021machine}.
%In the realm of multiple access techniques, the \ac{MPR} which is linked with interference management has emerged as a pivotal area of focus, addressing the increasingly complex challenges associated with spectrum utilization and mitigating interference to enhance the connectivity, network performance and user experience, toward the 6G era of machine-type communications \cite{Siddiqui2023,Miridakis2013,mahmood2021machine}.

The feasibility and increasing effectiveness of \ac{MPR} techniques, such as \ac{SIC}, are also driving the emergence of a new approach to multiple access, known as \ac{GFMA} \cite{Shahab2020grantfree,Choi2022grantfree}.
Traditional demand assignment techniques for multiple access are grant-based, i.e., they rely on signaling between the transmitting node and a central node, typically the base station in a cellular network, responsible for managing the radio resource through a scheduling function.
Signaling takes time and bandwidth, which is worth, if the duration of the subsequent data transfer and the amount of data transferred in a single interaction outweighs the cost of signaling.
However, traffic patterns typical of \ac{IoT} often result in signaling overhead that can easily dominate and degrade system performance \cite{Bockelmann2016mMTC_PHYMAC,Kouzayha2017_IoTsignalingOverhead}.
In contrast to demand assignment techniques, classic random multiple access is designed to facilitate the sharing of the communication channel among a large population of users.
These users transmit short messages at random times, resulting in a fluctuating number of active users over time that cannot be easily tracked.
The core function of random multiple access is contention, which typically involves sensing or other forms of indirect coordination among nodes competing to use the channel.
\ac{GFMA} aims at reducing the burden of coordination, essentially shifting the complexity of multiple access from the \ac{MAC} layer to the physical layer, where advanced signal processing algorithms are employed to support \ac{MPR}.

In this paper, we aim to understand the impact of \ac{SIC} and the achievable performance trade-offs for a system of nodes that generate update messages at random times (as driven by application-dependent events).
We do not propose a specific new \ac{SIC} algorithm: instead, we aim to characterize the impact that an idealized \ac{SIC} algorithm could achieve.
To this end, we define a general model of multiple access, consisting of $n$ non-saturated nodes that generate update messages sent to a central collecting node, equipped with a \ac{SIC} receiver.
The considered multiple access incorporates an adaptive grant-free scheme where system parameters are tuned to optimize the achieved sum-rate.
We formulate an analytical model for the system, using a mean-field approximation.
The proposed analytical model is highly accurate in predicting performance compared to simulations.
The model is then used to gain insights into performance metrics such as energy consumption. and \ac{AoI}, which is a well-established metric primarily relevant for update message traffic \cite{Yates2021,Abd2019_AoI}.

%We believe that this work opens up several paths for further research. 
%These include implementing practical algorithms to realize the promises assessed by the model.
%\textcolor{blue}{Defining distributed algorithms to adapt multiple access parameters, inspired by the insights gained from the presented mode is another research line we are pursuing.}

The rest of the paper is organized as follows.
The related work is explored further in \cref{sec:relwork}.
\cref{sec:model} introduces the analytical model, including assumptions, definitions, and notations.
The analysis of the model is presented in \cref{sec:analysis}, deferring some mathematical details to Appendices.
Expressions of several performance metrics are derived from the model in \cref{sec:perfmetrics}.
Numerical results are reported in \cref{sec:num_eval}, where the model is validated against simulations and used to investigate the performance metrics.
Finally, conclusions are drawn in \cref{sec:conclusions}.

% -------------- Section end marker --------------
%                _       _
%               ( )_    ( )
%    ___  _   _ | ,_)   | |__     __   _ __   __
%  /'___)( ) ( )| |     |  _ `\ /'__`\( '__)/'__`\
% ( (___ | (_) || |_    | | | |(  ___/| |  (  ___/
% `\____)`\___/'`\__)   (_) (_)`\____)(_)  `\____)
%
% -------------- Section end marker --------------

\section{Related works}
\label{sec:relwork}

A vast literature has been devoted to massive \ac{IoT} and specifically to issues revolving around multiple access techniques for such applications \cite{Liu20175G,Mahyar2017}.

The main characteristics of those scenarios are: (i) a large population of nodes that need to transmit relatively short messages, in a sporadic way, as opposed to few, high-rate users as in most current cellular applications; (ii) status update traffic is a key component, if not dominant, with respect to typical transactional or content transfer applications involving human users.
This paradigm shift calls for grant-free multiple-access schemes \cite{Shahab2020grantfree}.
% and new performance metrics on the other side, e.g., \ac{AoI}, value of information \cite{Qamar2023_AoI}.

Grant-free multiple access is a paradigm shift, where the core functionality is laid down in the physical layer, rather than as a \ac{MAC} protocol.
Two major research areas have witnessed an extensive contribution in this regards, namely \ac{NOMA} and \ac{SIC}.

\ac{NOMA} has emerged as a transformative technology in wireless communication, promising significant improvements in spectral efficiency, user capacity, and system throughput \cite{Mahyar2017,Ali2017,Islam2017}.
Our main focus is toward power domain \ac{NOMA} introduced in \cite{Ding2016}, which assigns different power levels to users based on their individual channel conditions.
It has been shown in \cite{Ding2016} that power domain \ac{NOMA} outperforms conventional \ac{OMA} schemes in terms of user capacity and system throughput.

Power domain \ac{NOMA} has been extensively studied in the context of 5G and beyond wireless networks, \cite{Liu2017} and \cite{Liu20175G} have explored the integration of \ac{NOMA} with \ac{MIMO} technology within large-scale heterogeneous networks.
Their study highlights the synergistic advantages of integrating \ac{NOMA} and \ac{MIMO}, particularly in improving coverage probability, spectral efficiency, accommodating a large number of users simultaneously and enhancing overall network energy efficiency.
Furthermore, \cite{Chen2018} highlighted the implementation of \ac{NOMA} in \ac{URLLC} for mission-critical applications, such as industrial automation and vehicular communication.
Their findings demonstrated the potential of NOMA to meet stringent reliability and latency requirements in \ac{URLLC} scenarios.

Recent research has emphasized the practical implementation and challenges of \ac{NOMA} in real-world wireless systems.
For instance, \cite{Benjebbour2015} conducted experimental evaluations of NOMA in a test bed environment to assess its performance under various conditions.
Their experiments offered valuable insights into the practical deployment of \ac{NOMA} and its impact on system performance.
Results showed that \ac{NOMA} yielded performance gains of over $30$\% compared to \ac{OFDMA}.
Hence, \ac{NOMA} has been demonstrated as a well established key element for addressing the challenges of next-generation networks toward massive connectivity, spectral efficiency and network performance.

A major role in the implementation of \ac{NOMA} is played by interference management algorithms and specifically \ac{SIC} \cite{Higuchi2015}.
\ac{SIC} algorithms play a pivotal role in decoding multiple user signals concurrently, harnessing NOMA's potential to enhance spectral efficiency and user capacity \cite{Chen2019SIC}.
Recent research has extensively explored the intricacies of SIC, with the aim of optimizing its efficiency and effectiveness in interference mitigation \cite{Ali2017}. 
For instance \cite{Abu2019} provides the theoretical framework for the maximized sum-rate subject to the quality of service requirements in the presence of \ac{SIC}.
Theoretical models have been crucial in elucidating the fundamental limits and performance bounds of \ac{SIC} \cite{Sen2013SIC,Zhang2014SIC,Wildemeersch2014}.
Moreover, recent research has explored integrating machine learning and deep learning methodologies to enhance SIC algorithms, improving their robustness and adaptability in real-world scenarios \cite{Aref2020,Van2022}.

Building on these advancements in \ac{NOMA} and signal processing techniques, there is a growing need for new multiple access techniques for next-generation communication systems.
Grant-free multiple access has garnered significant attention in recent years, particularly in the context of supporting sporadic traffic and massive \ac{IoT} networks, especially massive \ac{MTC} \cite{Shahab2020grantfree,Choi2022grantfree}.
\ac{GFMA} techniques aim to alleviate the overhead associated with traditional scheduled access methods by allowing devices to transmit data without explicit permission from the base station.
This approach is particularly beneficial for \ac{IoT} applications.
%with sporadic traffic patterns and a large number of devices, where the overhead of scheduling and coordination becomes prohibitive \cite{Shahab2020grantfree}.

\ac{GFMA} has been studied extensively \cite{Shahab2020grantfree,Amini2021lowlatency,Kang2023grantfree}, in the context of massive \ac{IoT} networks with stringent reliability and latency requirements.
In \cite{Kim2019grantfree}, an analysis of a sporadic \ac{IoT} traffic scenario utilizing the \ac{GFMA} protocol from a \ac{MAC} layer perspective is presented.
The study highlighted that \ac{GFMA} effectively supports low-latency transmissions of sporadic traffic generated by a large number of \ac{IoT} devices, depending on sufficient \ac{MPR} capability.

\ac{AoI} is a critical metric for assessing the freshness of information in communication systems, especially for real-time applications such as monitoring and control systems \cite{kaul2012real}.
In grant-free random access protocols, maintaining a low \ac{AoI} is challenging due to the sporadic nature of data transmissions and the potential for collisions among multiple devices.
Several studies have addressed the issue of \ac{AoI} in the context of grant-free random access protocols, few of which are highlighted below.
In \cite{Ding2024}, the authors investigated \ac{NOMA} assisted grant-free transmission designs using two different pre-configured \ac{SNR} levels based on target data rates, evaluating their impact on \ac{AoI} as a key performance metric.
The findings indicate significant potential for exploring grant-free schemes from both \ac{AoI} and energy perspectives.
In \cite{Yu2021}, the authors emphasized the benefits of reducing \ac{AoI} under a grant-free random access scheme for massive \ac{MIMO}.
Additionally, a study in \cite{Huang2023} examined grant-free massive access using frame-less ALOHA to minimize \ac{AoI}.

Despite all these studies and advancements, there remains a significant gap toward understanding the existing trade-offs in terms of energy and \ac{AoI}, particularly in environments characterized by high device density and varying traffic patterns.
The scope of our work is much broader as compared to existing literature, as our model exhibits both sporadic and bursty traffic due to the system built-in characteristics which are based on grant-free mechanism.
The insights gained in terms of throughput, energy, age of information along with grant-free mechanism, for different traffic scenarios are quite impressive, they surely open up several paths toward research on next grant-free multiple access techniques.
Our study emphasizes the role of SIC in grant-free mechanisms to ensure low \ac{AoI} and conserve energy, thereby enhancing the overall performance of \ac{IoT} and \ac{MTC} networks.

% ------------- Section end marker --------------
%                _       _
%               ( )_    ( )
%    ___  _   _ | ,_)   | |__     __   _ __   __
%  /'___)( ) ( )| |     |  _ `\ /'__`\( '__)/'__`\
% ( (___ | (_) || |_    | | | |(  ___/| |  (  ___/
% `\____)`\___/'`\__)   (_) (_)`\____)(_)  `\____)
%
% -------------- Section end marker --------------

% -------------- Section end marker --------------

\section{System Model}
\label{sec:model}

We consider a network of $n$ nodes, sending update messages to a sink, referred to as \ac{BS}.
The time axis is slotted.
A transmission attempt is made by backlogged nodes in each slot with a given probability.
More in-depth, let $Q(t)$ denote the number of backlogged nodes at the beginning of slot $t$.
If $Q(t) = k$, a backlogged node attempts transmission with probability $p_k$ and picks its modulation and coding scheme according to a target \ac{SNIR} $\gamma_k$ ($1 \le k \le n$).
Assuming an \ac{AWGN} communication channel, the target \ac{SNIR} $\gamma$ is tied to the achievable spectral efficiency $\eta$ (bits per symbol)  according to $\eta = \log_2(1+\gamma)$.
A packet is correctly decoded if its average \ac{SNIR} at the \ac{BS} is no less than $\gamma$.
We assume also that feasible values of $\gamma$ are upper limited to some value $\gamma_{\text{max}}$, related to maximum available transmission power and target coverage distance of the \ac{BS}.

Let $L$ be the length of the transmitted packets and $W$ be the channel bandwidth.
It is assumed that the slot size just fits a packet transmission time plus a fixed overhead time.
The slot size for a given target \ac{SNIR} $\gamma$ is:
\begin{equation}
\label{eq:slottimeduration}
T(\gamma) = T_{\text{oh}} + \frac{ L }{ W \log_2(1+\gamma) }
\end{equation}
where $T_{\text{oh}}$ is a fixed time accounting for slot overhead, that does not scale with the target \ac{SNIR} $\gamma$.

If $Q(t) = k$, we have $\gamma = \gamma_k$.
Hence, we define the duration of the slot time, given that at the beginning of the slot $k$ nodes are backlogged:
\begin{equation}
\label{eq:Tkexpression}
% T_k = T( \gamma_k ) = \frac{ L }{ W \log_2( 1 + \gamma_k ) } \, , \quad k = 1,\dots,n.
T_k = \begin{cases}
     T( \gamma_k ) = T_{\text{oh}} + \frac{ L }{ W \log_2( 1 + \gamma_k ) }  &  k = 1,\dots,n, \\
    T(0) = T_{\text{oh}}   &  k = 0.
\end{cases}
\end{equation}
In case no node is backlogged at the beginning of a slot, the duration of the slot reduces to $T_{\text{oh}}$.

New messages are generated at each node according to a Poisson process with mean rate $\lambda$.
Messages are generated in upper layers and passed down to the MAC layer entity.
Once a node MAC entity is engaged with contention/transmission of a message, it cannot be interrupted.
If a MAC entity is engaged in contention/transmission of a message, a new arriving message is dropped. 
It is shown in \cite{Baiocchi2021ITC33} that having no buffer at the MAC level is beneficial to \ac{AoI}, which is the relevant metric in the considered use case of update messages.
The reason is that keeping new messages extends the time that a node is backlogged and contends with others for channel resources. 
The resulting higher level of contention affects adversely \ac{AoI}, setting back the potential advantage of maintaining the latest new message in a buffer.

The path gain of the communication channel is modeled as $G = G_d G_f$, where $G_d$ is the deterministic gain accounting for distance between the transmitting node and the \ac{BS} and $G_f$ is the fading gain accounting for multi-path.
We assume that the deterministic path gain component does not vary over time, while the fading component is sampled independently slot by slot from a negative exponential \ac{PDF} with mean 1 (Rayleigh fading).
The transmission power level $P_{\text{tx}}$ is adjusted to compensate for the deterministic path gain component.
Therefore, the average received power level $P_0$ at \ac{BS} is set to a target value, so that the probability of failing to decode a packet sent by a single transmitting node is no more than $\epsilon$.
If there is a single transmitting node, using transmission power $P_{\text{tx}}$, decoding of the transmitted packet is successful, if the \ac{SNR} exceeds $\gamma$, i.e., if
\begin{equation}
\label{eq:singletxreq}
\frac{G_f G_d P_{\text{tx}} }{ P_N } \ge \gamma
\end{equation}
where $P_N$ is the background noise power level. 
Since the average received power level $G_d P_{\text{tx}}$ must match the prescribed level $P_0$, the requirement in \cref{eq:singletxreq} translates to:
\begin{equation}
\frac{ G_f P_0 }{ P_N } \ge \gamma \qquad \text{w.p. } 1-\epsilon
\end{equation}
Given that $G_f$ has a negative exponential \ac{PDF}, $P_0$ is set so that $\mathcal{P}( G_f P_0 / P_N \ge \gamma ) = e^{ - \gamma P_N/P_0} = 1-\epsilon$.
Hence, the target \ac{SNR} level $S_0$ at the receiving \ac{BS} is set as follows:
\begin{equation}
\label{eq:targetaveragerxSNR}
S_0 = \frac{ P_0 }{ P_N } = \frac{ \gamma }{ -\log(1-\epsilon) } = \frac{ \gamma }{ c }
\end{equation}
where we have introduced the constant $c = -\log(1-\epsilon)$.

We assume an ideal \ac{SIC} receiver.
Let $h$ packets be received simultaneously in the same slot and let $S_j, \, j = 1,\dots,h$ be their respective received power levels, normalized with respect to the background noise power level\footnote{Note that $S_j = G_{f,j} P_0/P_N = G_{f,j} S_0$, where $G_{f,j}$ is the fading path gain of the $j$-th user and $S_0$ is given in \cref{eq:targetaveragerxSNR}.}.
Let the $S_j$'s be ordered in descending order, i.e., $S_1 \ge S_2 \dots \ge S_h$ (ties are broken at random).
The \ac{SIC} receiver works as follows.
For each given $\ell \in \{1,\dots,h \}$, provided decoding of packets $1,\dots,\ell-1$ be successful, packet $\ell$ is decoded successfully if and only if the following inequality holds:
\begin{equation}
\label{eq:SICdecodingcondition}
    \frac{ S_\ell }{ 1 + \sum_{ r = \ell+1 }^{ h }{ S_r } } \ge \gamma
\end{equation}
Note that we assume perfect interference cancellation.
Hence, the residual interference is due only to signals weaker than the $\ell$-th one.

% -------------- Section end marker --------------
%                _       _
%               ( )_    ( )
%    ___  _   _ | ,_)   | |__     __   _ __   __
%  /'___)( ) ( )| |     |  _ `\ /'__`\( '__)/'__`\
% ( (___ | (_) || |_    | | | |(  ___/| |  (  ___/
% `\____)`\___/'`\__)   (_) (_)`\____)(_)  `\____)
%
% -------------- Section end marker --------------

\section{Model Analysis}
\label{sec:analysis}

The model analysis is carried out by considering the point of view of a tagged node, say node $i$. 
We drop the subscript denoting the tagged node unless required to avoid ambiguity.
If not stated explicitly, it is understood that each node-specific variable or quantity refers to the tagged node.

We assume that the number of nodes that are backlogged at the beginning of the slot $t$, $Q(t)$, is known.
If $Q(t) = k$, system parameters are adjusted as follows:
\begin{itemize}
	\item The target \ac{SNIR} is set to $\gamma_k$.
	\item The transmission probability is set to $p_k$.
\end{itemize}

The values $\gamma_k$ and $p_k$ are chosen as those values that maximize the sum rate $U_k(p,\gamma)$ with $k$ backlogged nodes.
\begin{equation}
\label{eq:sumratexpression}
U_k(p,\gamma) = \log_2(1+\gamma) \overline{D}_k(p,\gamma)
\end{equation}
where $\overline{D}_k(p,\gamma)$ is the mean number of packets correctly decoded in one time slot, given that $k$ nodes are backlogged at the beginning of the slot.
The sum-rate represents the achieved spectral efficiency of the multiple access channel in bits/s/Hz \cite{Razzaque2022}.
The function $\overline{D}_k(p,\gamma)$ is given by:
\begin{equation}
\label{eq:overlineDexpression}
\overline{D}_k(p,\gamma) = \sum_{ h = 0 }^{ k }{ m_h(\gamma) \binom{k}{h} p^k (1-p)^{k-h} } 
\end{equation}
where $m_h(\gamma)$ is the mean number of packets successfully decoded, given that $h$ nodes transmit in the same time slot.
The functions $m_h(\cdot)$, for $h = 1,\dots,n$, can be estimated numerically using \cref{eq:SICdecodingcondition} by means of ad-hoc simulations.

Let $p_k^*$ and $\gamma_k^*$ be the values of $p$ and $\gamma$ that maximize $U_k(p,\gamma)$ for a given $k$ and $c = -\log(1-\epsilon)$ (see \cref{eq:targetaveragerxSNR}).
It is found numerically that a good approximation of those values is:
\begin{equation}
\label{eq:pstark}
p^*_k = \begin{cases}
     \frac{ 1 }{ k } & \text{ for } 1 \le k < k_c, \\
     1 & \text{ for } k \ge k_c.
\end{cases}
\end{equation}
\begin{equation}
\label{eq:gammastark}
\gamma^*_k = \begin{cases}
    \gamma_{\text{max}} & \text{ for } 1 \le k < k_c, \\
     \frac{ 1 }{ a_\gamma k + b_\gamma } & \text{ for } k \ge k_c.
\end{cases}
\end{equation}
for suitable constants $k_c$, $a_\gamma$ and $b_\gamma$.
As a matter of example, for $\epsilon = 0.1$ and $\gamma_{\text{max}} = 31$, it is found that $k_c = 6$, $a_\gamma = 0.39$, and $b_\gamma = 0.78$.

% \Asmad{I have updated the values based on Asmad's estimated parameters, and the value of $k_c$ is updated considering the equation equality as it is now so it is consistent.}

\cref{eq:pstark,eq:gammastark} lend themselves to an intuitive explanation.
As long as few nodes are backlogged, the highest sum-rate is attained if a high target \ac{SNIR} is set and only one node transmits on average in a time slot.
On the contrary, when many nodes are backlogged, a higher sum-rate is achieved if all nodes are allowed to transmit, yet setting a small enough level of target \ac{SNIR}, i.e., inversely proportional to the number of backlogged nodes.
A theoretical underpinning of this insight is addressed in \cite{ASYMPTOTICPAPERARXIV}.

Let us define the probability distribution of the number of backlogged nodes seen by a tagged node:
\begin{equation}
\label{eq:qPDF}
q_k = \binom{n-1}{k} b^k (1-b)^{n-1-k} \, , \quad k = 0,1,\dots,n-1
\end{equation}
Here $q_k$ is the probability that $k$ nodes are backlogged, out of the $n-1$ nodes different from the tagged one.
The parameter $b$ is the probability that a node is backlogged at the beginning of a slot.
We will see in \cref{subsec:txProb} that $b$ is computed using a fixed point equation.

The expression in \cref{eq:qPDF} entails an independence assumption among node states.
In fact, the evolution of the number of backlogged nodes over time exhibits Markov chain dependence.
Assuming the number of backlogged nodes at the beginning of a slot is given by \cref{eq:qPDF} amounts to a \emph{mean-field approximation}.
It turns out that the accuracy of this approximation is excellent in the considered model setting, thus providing a simple, scalable and effective model.

\subsection{Slot time}
The slot time size is a random variable.
It takes the value $T_k, \, k = 0,1,\dots,n$, in case $k$ nodes are backlogged at the beginning of the slot time, as given in \cref{eq:Tkexpression}.
% The slot time size in case there is no backlogged node is denoted with $T_0$, to keep notation uniform.
% The value of $T_0$ is given and remains constant, and does not depend on packet length and target \ac{SNIR}.

The slot time \ac{PDF} seen by the tagged node depends on whether the tagged node is backlogged or not.
In the following, the Laplace transform of the slot time \ac{PDF} is derived in either case.

Let us start with the case when the tagged node is idle.
The slot time is denoted with $X$ in this case.
The slot time seen by the tagged node when it is idle, while $k$ of the other nodes are backlogged, is:
\begin{equation}
\label{eq:Xdefinition}
X = T_k \quad  \text{w.p. } q_k, \quad k = 0,1,\dots,n-1
\end{equation}
The Laplace transform of the \ac{PDF} of $X$ is:
\begin{equation}
\label{eq:LTPDFXdefine}
\varphi_{X}(s) = \sum_{ k = 0 }^{ n-1 }{ q_k e^{ - s T_k } }
\end{equation}

Let $X^\prime$ denote the slot time when the tagged node is backlogged.
The expressions of \ac{PDF} of $X^\prime$ and its Laplace transform are similar to those of $X$, except that now there are $k+1$ backlogged nodes with probability $q_k$.
Hence:
\begin{equation}
\label{eq:Xprimedefinition}
X^\prime = T_{k+1} \quad  \text{w.p. } q_k, \quad k = 0,1,\dots,n-1
\end{equation}
and
\begin{equation}
\label{eq:LTPDFXprimedefine}
\varphi_{X^\prime}(s) = \sum_{ k = 0 }^{ n-1 }{ q_k e^{ - s T_{k+1} } }
\end{equation}

\subsection{Inter-departure time}
\label{subsec:Ytime}
\begin{figure}[t]
\centering
\includegraphics[width=0.8\columnwidth]{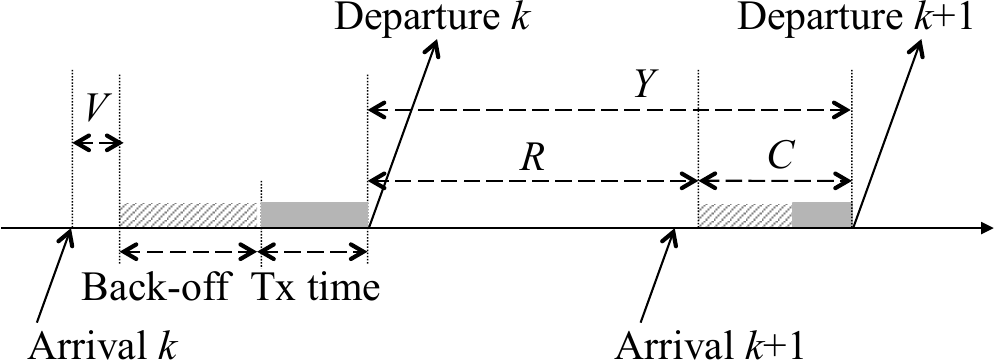}
\caption{Definition of the inter-departure time $Y$, the idle time between a departure and the end of the slot where the subsequent arrival occurs, $R$, and the time, where the node is busy contending and eventually transmitting, $C$. The dashed area highlights the time slot(s) where the node is backlogged, but backing off. The dark  shaded area highlight the time slot where the node transmits.}
\label{fig:definitionofYRC}
\end{figure}

Let $Y$ denote the inter-departure time between two consecutive packets transmitted by the tagged node.
Let also $C$ denote the contention time, defined as the sum of slot times it takes for the tagged node to transmit, once it becomes backlogged.
Let $R$ denote the time elapsing since the end of a transmission of the tagged node, until the end of the slot where it becomes backlogged again.
The relationship among $Y$, $R$ and $C$ is illustrated in \cref{fig:definitionofYRC}.

Then we have:
\begin{equation}
\label{eq:Ydefinition}
Y = R + C = \sum_{ i = 1 }^{ N }{ X(i) } + \sum_{ i = 1 }^{ M }{ X^\prime(i) }
\end{equation}
where $N$ and $M$ are two discrete random variables, representing the number of slots that the tagged node stays idle and the number of slots that it is contending and eventually making its transmission, respectively.

Thanks to the independence assumption among states of nodes, hence among slot times, from \cref{eq:Ydefinition}, we have:
\begin{equation}
\varphi_Y(s) = \mathrm{E}[ e^{ - s Y } ] = \varphi_R(s) \varphi_C(s)
\end{equation}

In Appendix \ref{app:A} it is shown that the expressions of the Laplace transforms of $C$ and $R$ are given by:
\begin{equation}
\label{eq:LTofPDFofC}
\varphi_{C}(s) = \frac{ \sum_{ k = 0 }^{ n-1 }{ q_k p_{k+1} e^{ - s T_{k+1} } } }{ 1 - \sum_{ k = 0 }^{ n-1 }{ q_k (1-p_{k+1}) e^{ - s T_{k+1} } } }
\end{equation}
and
\begin{equation}
\label{eq:LTofPDFofR}
\varphi_{R}(s) = \frac{ \varphi_X(s) - \varphi_X(s+\lambda) }{ 1 - \varphi_X(s+\lambda) }.
\end{equation}
Expressions of moments of $C$ and $R$ are given in Appendix \ref{app:A} as well.

\subsection{Probability of being backlogged}
\label{subsec:txProb}

In order to find $b$, the probability that a node is backlogged at the beginning of a slot, we use the renewal reward theorem, yielding:
\begin{equation}
\label{eq:rewardtheoremforb}
b = \frac{ \mathrm{E}[ M ] }{  \mathrm{E}[ N ] +  \mathrm{E}[ M ] }
\end{equation}
where $M$ and $N$ are defined in \cref{subsec:Ytime}.
The probability distributions of $N$ and $M$ are derived in Appendix \ref{app:B}, yielding:
\begin{equation}
\mathcal{P}( N = j ) = \left[ 1 - \varphi_X(\lambda)  \right] \left[ \varphi_X(\lambda) \right]^{j-1} \, , \; j \ge 1.
\end{equation}
\begin{equation}
\mathcal{P}( M = j ) = \overline{p}^\prime \left( 1 - \overline{p}^\prime \right)^{j-1} \, , \quad j \ge 1
\end{equation}
where $\overline{p}^\prime = \sum_{ k = 0 }^{ n-1 }{ q_k p_{k+1} }$.

Taking the mean values of $M$ and $N$ and using \cref{eq:rewardtheoremforb}, it follows that:
\begin{equation}
\label{eq:bfixedpoint}
b = \frac{ 1/\overline{p}^\prime }{ \frac{ 1 }{ 1 - \varphi_X(\lambda) } + 1/\overline{p}^\prime } = \frac{ 1 }{ 1 + \overline{p}^\prime / [ 1 - \varphi_X(\lambda) ] }
\end{equation}

It can be checked that $b \rightarrow 0$ as $\lambda \rightarrow 0$ (light traffic regime), while $b \rightarrow  1 / ( 1 + \overline{p}^\prime )$ for $\lambda \rightarrow \infty$ (heavy traffic regime)\footnote{
The backlogged probability $b$ cannot attain 1, since it is assumed that nodes have no buffer.
Hence, a node that completes a transmission attempt must wait at least one slot time, before a new message is handed out from the upper layers to the MAC layer of that node.
}.

The probability $b$ is a function of the $q_k$'s through $\overline{p}^\prime$ and $\varphi_X(\lambda)$.
In turn, the $q_k$'s depend on $b$.
Hence $b$ is computed by solving \cref{eq:bfixedpoint} as a fixed point equation $b = F(b)$ for $b \in [0,1]$.
Since $F(\cdot)$ defines a continuous map of the compact interval $[0,1]$ onto itself, we can appeal to Brouwer's theorem to guarantee that a fixed point exists for $b \in [0,1]$.
The proof of uniqueness of the fixed point of \cref{eq:bfixedpoint} is given in Appendix \ref{app:C}.

%  ___  _   _ | ,_)   | |__     __   _ __   __
%  /'___)( ) ( )| |     |  _ `\ /'__`\( '__)/'__`\
% ( (___ | (_) || |_    | | | |(  ___/| |  (  ___/
% `\____)`\___/'`\__)   (_) (_)`\____)(_)  `\____)
%
% -------------- Section end marker --------------

\section{Performance metrics}
\label{sec:perfmetrics}

Analytical expressions of all considered performance metrics (success probability, throughput, \ac{CBR}, \ac{AD}, \ac{AoI}, energy consumption) are derived in this Section.

\subsection{Successful delivery probability, throughput and \ac{CBR}}
\label{subsec:throughput}

By renewal reward theorem, the success probability can be written as the ratio of the mean number of correctly decoded packets per slot divided by the mean number of transmitted packets per slot.
Let $J_D$ and $J_T$ be defined as the number of packets correctly decoded and transmitted per slot, respectively.
Then, conditioning on the number of backlogged nodes, we have:
\begin{equation}
%\label{ }
\mathrm{E}[ J_D | Q = k ] = \sum_{ h = 1 }^{ k }{ m_h(\gamma_k) \binom{k}{h} p_k^h (1-p_k)^{k-h} }
\end{equation}
and
\begin{equation}
%\label{ }
\mathrm{E}[ J_T | Q = k ] = k p_k
\end{equation}
Let $w_k$ be the probability that $k$ nodes are backlogged::
\begin{equation}
%\label{ }
w_k = \binom{n}{k} b^k (1-b)^{n-k} \, , \quad k = 0,\dots,n,
\end{equation}
Then the probability of success $P_s$ is given by:
\begin{equation}
\label{eq:Psformula}
P_s = \frac{ \sum_{ k = 1 }^{ n }{ w_k \mathrm{E}[ J_D | Q = k ] } }{ \sum_{ k = 1 }^{ n }{ w_k \mathrm{E}[ J_T | Q = k ] } }
\end{equation}

Given the generation rate $\lambda$ of messages at a node, there are two sources of message loss: (i) dropping of arriving messages when the tagged node is busy in contention or transmission; (ii) failed decoding.
The mean rate of messages sent on air by a node is $1/\mathrm{E}[Y]$, i.e., the mean inter-departure rate.
The net throughput in messages per unit time is therefore:
\begin{equation}
\Theta = \frac{ P_s }{ \mathrm{E}[Y] } 
\end{equation}
The normalized throughput is given by $\Theta_{\text{norm}} = \Theta/\lambda$.
The throughput in $\si{\bit\per\second}$ can be obtained by considering the message length $L$, i.e., it is $\Theta_{\text{bps}} =L \Theta$.

The \ac{CBR} is the mean fraction of time that the \ac{BS} senses the channel as busy.
It can be evaluated as follows\footnote{Here and in the following, we let $p_0 = 0$ for ease of notation. A transmission probability value is not actually needed in case there is no backlogged node.}:
\begin{equation}
\text{CBR} = 1 - \frac{ \sum_{ k = 0 }^{ n }{ w_k ( 1 - p_k )^k T_k } }{ \sum_{ k = 0 }^{ n }{ w_k T_k } }
\end{equation}
In this expression it is assumed that the slot time size is a function of the number of \emph{backlogged} nodes, independently of how many of them actually transmit.

% \AB{Attenzione. Stiamo supponendo che la durata dello slot dipenda dal numero di utenti backlogged ANCHE SE NESSUNO trasmette.
% In alternativa, potremmo dire che la durata e' $T_0$ w.p. $\sum_{ k = 0 }^{ n }{ q_k (1-p_k)^k }$.}

\subsection{Access delay}
\label{subsec:accessdelay}

\begin{figure}[t]
\centering
\includegraphics[width=0.5\columnwidth]{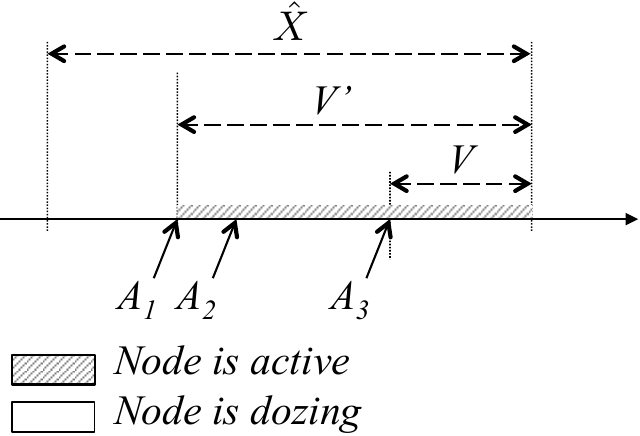}
\caption{Definition of the time slot where an arrival occurs, $\hat{X}$, the time span since the \emph{first} arrival in that slot (denoted with $A_1$) until the end of that slot, $V^\prime$, and the time span since the \emph{last} arrival in that slot (denoted with $A_3$) until the end of that slot, $V$. The shaded area highlights the portion of the slot time where the tagged node is active, hence consuming energy.}
\label{fig:definitionofXhatVVprime}
\end{figure}

The access delay $D$ is defined as the interval between the arrival time of a message and the completion of the transmission time of that message.
The access delay is $D = V+C$, where $V$ is the time elapsing since the \emph{last} arrival within a time slot and the end of that time slot.
The Laplace transform of the \ac{PDF} of $D$ is:
\begin{equation}
\varphi_D(s) = \varphi_C(s) \varphi_V(s) 
\end{equation}
where $\varphi_C(s)$ is given in \cref{eq:LTofPDFofC}.
There remains to evaluate the Laplace transform of the \ac{PDF} of $V$.
As usual, $X$ denotes the slot time when the tagged node is idle.
Let also $\hat{X}$ denote the time interval $X$, conditional on at least one arrival occurring at the tagged node in that slot time. 
The relationship among arrivals and the time spans $\hat{X}$ and $V$ is depicted in \cref{fig:definitionofXhatVVprime}.

The \ac{CCDF} of $\hat{X}$ is found as follows::
\begin{align*}
\mathcal{P}( \hat{X} > x ) &= \mathcal{P}( X > x | A(0,X) > 0 )  \\
  &= \frac{ \int_{ x }^{ \infty }{ (1-e^{ - \lambda u } ) f_X(u) \, du } }{ \int_{0}^{\infty}{ (1-e^{ - \lambda u } ) f_X(u) \, du } }  \\
  &= \frac{ \int_{ x }^{ \infty }{ (1-e^{ - \lambda u } ) f_X(u) \, du } }{ 1 - \varphi_X(\lambda) } 
\end{align*}
where $A(u,v)$ is the number of arrivals in the time interval $[u,v]$ according to the Poisson process with mean rate $\lambda$.
Taking the derivative, we obtain the \ac{PDF} of $\hat{X}$:
\begin{equation}
f_{\hat{X}}(x) =  \frac{ (1-e^{ - \lambda x } ) f_X(x) }{ 1 - \varphi_X(\lambda) } 
\end{equation}
%The Laplace transform of the \ac{PDF} of $\hat{X}$ is:
%\begin{equation}
%\varphi_{\hat{X}}(s) = \frac{ \varphi_X(s) - \varphi_X(s+\lambda) }{  1 - \varphi_X(\lambda) } 
%\end{equation}
%By deriving, the mean of $\hat{X}$ is easily found to be:
%\begin{equation}
%\label{eq:meanofhatX}
%\mathrm{E}[ \hat{X} ] =  \frac{ \mathrm{E}[ X ] }{ 1 - \varphi_X( \lambda ) } + \frac{ \varphi_X^\prime(\lambda) }{ 1 - \varphi_X(\lambda) }
%\end{equation}

We are now ready to derive the \ac{CCDF} of $V$:
\begin{equation}
\label{eq:CCDFV}
\mathcal{P}( V > t ) = \int_{t}^{\infty}{ \mathcal{P}( V > t | \hat{X} = x ) f_{\hat{X}}(x) \, dx }
\end{equation}
Let $P(t|x) =  \mathcal{P}( V > t | \hat{X} = x )$ denote the conditional probability inside the integral.
It can be evaluated as follows:
\begin{align*}
P(t|x) &= \frac{ \mathcal{P}( A(0,x-t) > 0, A(x-t,x) = 0 ) }{ \mathcal{P}( A(0,x) > 0 ) }   \\
  &= \frac{ (1-e^{ -\lambda (x-t) } ) e^{ -\lambda t } }{ 1-e^{ - \lambda x} } = \frac{ e^{ -\lambda t } - e^{ -\lambda x } }{ 1 - e^{ -\lambda x } }
\end{align*}
Substituting the expression of the \ac{PDF} of $\hat{X}$ and the result above into \cref{eq:CCDFV}, we  get:
\begin{equation}
\mathcal{P}( V > t ) = \int_{t}^{\infty}{ \left( e^{ -\lambda t } - e^{ -\lambda x } \right) \frac{ f_X(x) }{ 1 - \varphi_X(\lambda) } \, dx}
\end{equation}
Taking the derivative, we finally find the \ac{PDF} of $V$:
\begin{equation}
f_V(t) = \frac{ \lambda e^{ -\lambda t } \mathcal{P}( X > t ) }{ 1 - \varphi_X(\lambda) }
\end{equation}

The corresponding Laplace transform is:
\begin{equation}
\label{eq:LTofPDFofV}
\varphi_V(s) = \frac{ \lambda }{ 1 - \varphi_X(\lambda) } \frac{ 1 - \varphi_X(s+\lambda) }{ s + \lambda }
\end{equation}

Taking the mean of $D$ as the sum of the means of $C$ and $V$, which are in turn calculated from the Laplace transforms of their \acp{PDF} in \cref{eq:LTofPDFofC,eq:LTofPDFofV} respectively, we find:
\begin{equation}
%\label{ }
\mathrm{E}[D] = \frac{ \overline{T}^\prime }{ \overline{p}^\prime } + \frac{ 1 }{ \lambda } + \frac{ \varphi_X^\prime(\lambda) }{ 1 - \varphi_X(\lambda) }
\end{equation}
where $\overline{T}^\prime = \sum_{ k = 0 }^{ n - 1 }{ q_k T_{k+1 } }$, $\overline{p}^\prime = \sum_{ k = 0 }^{ n - 1 }{ q_k p_{k+1 } }$ and $\varphi_X(s)$ is given in \cref{eq:LTPDFXdefine}.

\subsection{Age of Information}
\label{subsec:AoI}

\ac{AoI} is the age of the messages received at the \ac{BS} from the tagged node.
When a message is transmitted, it has already accumulated an age corresponding to its access time $D$.
We account also for this initial delay in the evaluation of the \ac{AoI}.

The \ac{AoI} $H$ is akin to the excess time in a renewal process.
Its \ac{CDF}, given $D = u$, is:
\begin{equation}
\label{eq:CDFofHconditionalonD}
\mathcal{P}( H \le x | D = u ) = \begin{cases}
    \int_{ 0 }^{ x-u }{ \frac{ G_Z(t) }{ \mathrm{E}[Z] } \, dt }  & x \ge u, \\
    0  & \text{otherwise}.
\end{cases}
\end{equation}
Here $Z$ is the time between the reception at the \ac{BS} of two consecutive \emph{successful} messages from the tagged node.
It is therefore the sum of consecutive inter-departure times, those elapsing between two consecutive transmissions of the tagged node that do not run into a decoding failure.
We have:
\begin{equation}
\label{eq:Zdef}
    Z = \sum_{ i = 1 }^{ K }{ Y(i) }
\end{equation}
where $\mathcal{P}( K = j ) = (1-P_s)^{j-1} P_s, \, j \ge 1$, with $P_s$ given in \cref{eq:Psformula}.
Thanks to the independence of the $Y(i)$'s and $K$, the Laplace transform of the \ac{PDF} of $Z$ is given by:
\begin{equation}
\label{eq:LTofPDFofZ}
\varphi_Z(s) = \frac{ P_s \varphi_Y(s) }{ 1-(1-P_s)\varphi_Y(s) }
\end{equation}

Note that $D$ is independent of $Z$ under the mean field approximation.
Removing the conditioning in \cref{eq:CDFofHconditionalonD}, we get:
\begin{equation}
F_H(x) = \mathcal{P}( H \le x ) = \int_{ 0 }^{ x }{ f_D(u) \, du  \int_{ 0 }^{ x-u }{ \frac{ G_Z(t) }{ \mathrm{E}[Z] } \, dt } }
\end{equation}
The \ac{PDF} of $H$ is found by deriving $F_H(x)$, thus getting:
\begin{equation}
%\label{ }
f_H(x) = \frac{ 1 }{ \mathrm{E}[Z] } \, \int_{ 0 }^{ x }{ f_D(u) G_Z(x-u) \, du }
\end{equation}

The Laplace transform of the \ac{PDF} of $H$ is:
\begin{equation}
\label{eq:LTofPDFofH}
\varphi_H(s) = \varphi_D(s) \frac{ 1 - \varphi_Z(s) }{ s \mathrm{E}[Z] }
\end{equation}

Since it is $\mathrm{E}[Z] = \mathrm{E}[Y]/P_s$, using \cref{eq:LTofPDFofH,eq:LTofPDFofZ} we have:
\begin{equation}
\label{eq:LaptrasfPDFH}
\varphi_H(s) = \varphi_D(s) \frac{ P_s \left[ 1 - \varphi_Y(s) \right] }{ s \mathrm{E}[Y] \left[ 1 - (1-P_s) \varphi_Y(s) \right] }
\end{equation}
where $\varphi_Y(s)$ is found in \cref{subsec:Ytime}.

The mean \ac{AoI} is calculated by deriving \cref{eq:LaptrasfPDFH} and setting $s = 0$:
\begin{equation}
\mathrm{E}[H] = \mathrm{E}[D] + \frac{ \mathrm{E}[ Y^2 ] }{ 2 \mathrm{E}[Y] } + \mathrm{E}[ Y ] \left( \frac{ 1 }{ P_s } - 1 \right)
\end{equation}
with:
\begin{align}
	&\mathrm{E}[ Y ] = \mathrm{E}[ C ]+ \mathrm{E}[ R ]  \\
	&\mathrm{E}[ Y^2 ] = \mathrm{E}[ C^2 ] + 2 \mathrm{E}[ C ] \mathrm{E}[ R ] + \mathrm{E}[ R^2 ]
\end{align}

We can find an asymptotic approximation of the \ac{CCDF} of the \ac{AoI}.
The dominant pole of the Laplace transform of \ac{PDF} of \ac{AoI} is $-\zeta$, to be found as the smallest modulus root of the equation $\varphi_Y(-\zeta) = 1/(1-P_s)$.
The \ac{CCDF} of \ac{AoI} can be approximated as a shifted exponential distribution, i.e.,
\begin{equation}
G_{H}(t) \approx \min\{ 1, a e^{-\zeta t} \}
\end{equation}
The coefficient $a$ can be found by imposing the mean value of this approximation to be the same as the exact one, $ \mathrm{E}[H]$.
Since the mean of a non-negative random variable with \ac{CCDF} given by $G(t)$ equals $\int_{0}^{\infty}{G(t)dt}$, it is easy to find that it must be $a = \exp(\zeta \mathrm{E}[H] -1)$.
Then, we have:
\begin{equation}
G_{H}(t) \approx \min\{ 1, e^{-\zeta ( t - \mathrm{E}[H] )-1} \}
\end{equation}

\subsection{Energy consumption}
\label{subsec:energy}

Let $P_{\text{d}}$ and $P_{\text{a}}$ denote respectively the power consumed by a node when it is idle (dozing) and the power consumed when the node is active.
On top of $P_{\text{a}}$, the node also consumes the transmission power, when it transmits.
For typical device settings, $P_{\text{d}}$ is a small fraction of $P_{\text{a}}$, while $P_{\text{a}}$ is comparable to $P_{\text{a}} + P_{\text{tx}}$.
The average energy consumed by a node during an inter-departure time is given by:
\begin{equation}
\label{eq:averageenergyperinterdeptime}
E_{\text{d}} = P_{\text{d}} ( \mathrm{E}[ R ] - \mathrm{E}[ V^\prime ] ) + P_{\text{a}} ( \mathrm{E}[ V^\prime ] + \mathrm{E}[ C ] ) + E_{\text{tx}}   % P_{\text{tx}} \mathrm{E}[ X^\prime ]
% P_{\text{d}} (R-V^\prime) + P_{\text{a}} (V^\prime+C) + P_{\text{tx}} X^\prime
\end{equation}
where $V^\prime$ is the time elapsing since the \emph{first} arrival of a message in a slot time when the tagged node is idle and the end of that slot time, and $E_{\text{tx}}$ is the mean energy spent by a node for a transmission.

The reason why $V^\prime$ appears in this calculation instead of $V$ is that the MAC entity of the node activates (therefore consuming $P_{\text{a}}$) as soon as the first message is pushed down to the MAC layer from upper layers.
Thanks to the time-reversibility of the Poisson process, it is $V^\prime = \hat{X} - V$, where we recall that $\hat{X}$ is the slot time conditional on at least one arrival occurring.
The mean of $V^\prime$ is found by exploiting the time-reversibility of Poisson arrivals and using \cref{eq:LTofPDFofV} to derive the mean of $V$:
\begin{equation}
\mathrm{E}[ V^\prime ] = \mathrm{E}[ \hat{X} ] - \mathrm{E}[ V ] = \frac{ \mathrm{E}[ X ] }{ 1 - \varphi_X( \lambda ) } - \frac{ 1 }{ \lambda }
\end{equation}

Let us now turn to the evaluation of $E_{\text{tx}}$.
Because of power control, a transmitting node sets its transmission power as a function of the target \ac{SNIR}.
More in depth, let us consider a tagged node and assume it is backlogged.
Given that $k$ nodes are backlogged, other than the tagged one, the target \ac{SNIR} is set to $\gamma_{k+1}$ and the slot time is $T_{k+1}$ (see \cref{sec:model}).
Then, the tagged node sets its transmission power $P_{\text{tx},k+1}$ so that $G_d(r) P_{\text{tx},k+1} / P_N = \gamma_{k+1}/c$, where $G_d(r)$ is the deterministic path loss at distance $r$.
Hence $P_{\text{tx},k+1} = P_N \gamma_{k+1} /  (c \, G_d(r) )$.
Averaging over all nodes, assuming uniform node scattering over a circle of radius $R$ centered at the \ac{BS}, we have:
\begin{equation}
\label{eq:settingofPtxkp1}
\overline{P}_{\text{tx},k+1} = \mathrm{E}[ P_{\text{tx},k+1} ] = P_N \frac{ \gamma_{k+1} }{ c } \int_{0}^{R}{ \frac{ 1 }{ G_d(r) } \frac{ 2 r }{ R^2 } \, dr }
\end{equation}
for $k = 0,\dots,n-1$.
The average energy used for transmission is therefore:
\begin{equation}
\label{eq:defEtx}
E_{\text{tx}} = \sum_{ k = 0 }^{ n - 1 }{ q_k \overline{P}_{\text{tx},k+1} T_{k+1} }
\end{equation}

The two-ray ground path loss model described in \cite{Sommer2012TRG} is assumed for the deterministic gain component $G_d$.
Antenna heights are set to 1 m and 4 m for the transmitting nodes and the receiving \ac{BS}, respectively.
Then, the integral in \cref{eq:settingofPtxkp1} can be evaluated numerically, once a radius $R$ is fixed.

There remains to set a range $R$ within which nodes are scattered around the \ac{BS}.
This is done by requiring that any node is able to match the power control requirement even in the worst case.
The worst case corresponds to a node at distance $R$ that has to meet a target \ac{SNIR} of $\gamma_{\text{max}}$.
In that case, we require that $G_d(R) P_{\text{tx,max}} / P_N \ge \gamma_{\text{max}} / c$, so that $R$ is the largest value that meets the following inequality:
\begin{equation}
\frac{ G_d(R) P_{\text{tx,max}} }{ P_N } \ge \frac{ \gamma_{\text{max}} }{ c }
\end{equation}
This is found numerically, by using the path loss model in \cite{Sommer2012TRG} for $G_d(r)$.

As for the energy consumed in message generation, let $E_{\text{g}}$ be the fixed amount of energy required to generate a new update message.
The mean number of messages generated during an inter-departure time is $\lambda \mathrm{E}[ Y ]$. 
Hence, the mean amount of energy used to generate new messages for one message transmission attempt is given by $E_{\text{g}} \lambda \mathrm{E}[ Y ]$.

As reference energy metric, we define the average energy spent per delivered packet.
It is the sum of the energy spent for dozing until the arrival of new messages, message generation, contention and transmission.
Accounting for the mean number of transmission attempts required to deliver a successful message, namely $1/P_s$, the mean energy consumed per delivered packet is given by:
\begin{equation}
\label{eq:meanenergyperdeliveredpkt}
\overline{E} = \frac{ E_{\text{g}} \lambda \mathrm{E}[ Y ] + E_{\text{d}} }{  P_s }  
\end{equation}
where $E_{\text{d}}$ is given in \cref{eq:averageenergyperinterdeptime}.

% -------------- Section end marker --------------
%                _       _
%               ( )_    ( )
%    ___  _   _ | ,_)   | |__     __   _ __   __
%  /'___)( ) ( )| |     |  _ `\ /'__`\( '__)/'__`\
% ( (___ | (_) || |_    | | | |(  ___/| |  (  ___/
% `\____)`\___/'`\__)   (_) (_)`\____)(_)  `\____)
%
% -------------- Section end marker --------------

\section{Model Validation and Performance Evaluation}
\label{sec:num_eval}

To validate the analytical model an ad-hoc simulation code has been implemented in \matlab{}.
The simulation code  fully considers the evolution of the system.
More in-depth, each node evolves according to a two-state Markov chain.
A node is idle until a new message is generated.
At the end of the slot where a new message is generated, the node transitions to the active state, where it contends for the channel.
While active, it transmits with probability $p_k$ in a slot, if $k$ nodes are backlogged at the beginning of that slot.
Immediately after having transmitted, the node moves back to the idle state.

The slot time in the simulation is set to $T_k$ if $k$ nodes are backlogged at the beginning of the slot.
The time-varying size of slot times gives rise to a complex interplay between nodes, since the more nodes are active in one slot, the longer its duration, the higher the probability that a new message arrives at those nodes that are not active in that slot.

The validation of the model against simulations highlights that the analytical model matches simulations with high accuracy, despite being based on the mean-field approximation.
This gives evidence of its usefulness as an effective tool for the evaluation and dimensioning of the considered system.

Simulations have been run for several different scenarios to test the accuracy of the analytical model.
Here we provide a sample of numerical results.
In the results presented, we consistently set the number of nodes to $n = 50$, and the packet size to $L = \SI{500}{\byte}$, while varying the mean message generation time $S = 1/\lambda$ between 1 ms and 1000 ms.
% \SIrange{1}{1000}{\milli\second}.
The numerical values of the main system parameters are listed in Table \ref{tab:1}.
With those numerical values, the coverage range $R$ in \cref{subsec:energy} is $R \approx 876$ m.
We emphasize that all power levels and propagation parameters are needed only for the evaluation of the energy-related metric.

Simulations are displayed along with the $95$\% confidence intervals.
In all plots, results from the analytical model are shown as line curves, while simulation results are displayed as square markers.

\begin{table}[t]
\centering
\caption{Parameter values used for numerical evaluation.}
    \begin{tabular}{c c c c}
    \hline
    Parameter & Value & Parameter & Value \\
    \hline
    $P_N$  & $-107$ dBm & $W$ & $1$ MHz   \\
    $P_{\text{a}}$ & 1 mW & $L$ & $500$ bytes \\
    $P_{\text{d}}$ & 10 $\mu$W & $n$ & $50$ \\
    $P_{\text{tx,max}}$ & 100 mW & $\epsilon$ & 0.1  \\
    $E_{\text{g}}$ & 10 $\mu$J & $T_{\text{oh}}$ & 1 ms  \\
    $k_c$ & 6 & $\gamma_{\text{max}}$ & 31  \\
    $a_\gamma$ & 0.39 & $b_\gamma$ & 0.78  \\
    \hline
    \end{tabular}
\label{tab:1}
\end{table}

\subsection{Light and heavy traffic regimes}
\label{subsec:lightheavyregimes}

Numerical results are plotted as a function of the mean message generation time, $S = 1/\lambda$.
We consider a quite stretched range of $S$ values, to highlight the existence of two different operational regimes of the system.
Low values of $S$, lying  on the left side of the $x$-axis of each plot, correspond to nodes generating new messages very frequently, i.e., with mean generation time smaller than the average slot size.
We refer to this region as a heavy traffic regime.
On the opposite side of $x$-axis, large values of $S$ correspond to the light traffic regime, where nodes generate new messages infrequently, imposing a light load on the channel, given that message generation times are much bigger than the average slot size.

Under heavy traffic regimes, it is $k \gg 1$, hence $p_k = 1$ with high probability.
Backlogged nodes transmit simultaneously and packets are decoded exploiting essentially \ac{SIC}.
In the light traffic regime, where $k < k_c$ with high probability, hence $p_k = 1/k$, MAC regulation takes precedence, resulting in backlogged nodes transmitting one at a time with a high probability.

The transition region between the two regimes is identified by a critical mean generation time $S_\infty$, or equivalently, a critical rate $\lambda_\infty = 1/S_\infty$.

The critical message generation rate $\lambda_\infty$ is identified as follows.
The achievable message throughput rate per node given that $k$ nodes are backlogged is:
\begin{equation}
\label{eq:lambdacrit}
\mu_k = \frac{ W U_k(p^*_k,\gamma^*_k) }{ n L }
\end{equation}
where $U_k(\cdot,\cdot)$ is defined in \cref{eq:sumratexpression}, $p^*_k$ and $\gamma^*_k$ are specified in \cref{eq:pstark,eq:gammastark}.

The upper bound of the sustainable message throughput rate\footnote{
Note that any message generation rate is ``sustainable'' for the analyzed multiple access system, since nodes discard new messages when they are busy contending for the channel or transmitting. So, there is no queue instability issue at each node. Moreover, the number of nodes is finite. The analysis carried out here to identify the critical threshold separating the light and heavy traffic regimes relaxes the system model, assuming an infinite number of nodes.}
 is given by $\lambda_\infty = \lim_{ k \rightarrow \infty }{ \mu_k }$, if this limit exists.

Numerical evaluation of $\overline{D}( p^*_k,\gamma^*_k )$ for a wide range of values of $k$ points out the following asymptotic behavior as $k$ grows (see \cite{ASYMPTOTICPAPERARXIV} for a theoretical assessment of this result):
\begin{equation}
    \overline{D}( p^*_k,\gamma^*_k ) \sim a_D k, \quad k \rightarrow \infty
\end{equation}
for a suitable constant $a_D > 0$.
With numerical values listed in \cref{tab:1}, it is $a_D \approx 0.89$.
Putting together this result with the expression of $\gamma^*_k$ provided in \cref{eq:gammastark} for $k \ge k_c$, it follows that:
\begin{align}
    U_k(p^*_k,\gamma^*_k) &= \log_2(1+\gamma^*_k) \overline{D}( p^*_k, \gamma^*_k ) \sim \frac{ a_D k }{ ( a_\gamma k + b_\gamma ) \log 2 }   \nonumber \\
    &\sim \frac{ a_D }{ a_\gamma \log 2 } = U_\infty, \quad k \rightarrow \infty
\end{align}
With $\epsilon = 0.1$ and $\gamma_{\text{max}} = 31$, we have $U_\infty \approx 2.99$ bit/s/Hz.
Using \cref{eq:lambdacrit}, we end up defining the following critical message generation rate:
\begin{equation}
    \lambda_\infty = \frac{ W U_\infty }{ n L }
\end{equation}
With the numerical values in \cref{tab:1}, it is $S_\infty = 1/\lambda_\infty \approx 67$ ms.

The critical rate $\lambda_\infty$ gives the upper bound of the sustainable message generation rate in case \emph{all messages must be delivered successfully} and there is a potentially unbounded number of nodes.
These are not the operational conditions of the considered system model, where instability is kept off by the limited number of admitted nodes and by the message handling policy of nodes.
Yet, we will see that the critical rate $\lambda_\infty$ marks the turning point, where the working regime of the system steers from heavy to light traffic regime, as the mean generation time grows.
In the ensuing plots, when $S$ is displayed on the $x$-axis, the critical message generation rate $S_\infty = 1/\lambda_\infty$ is marked by a vertical dashed line.

\subsection{Optimal target \ac{SNIR} and transmission probability}
\label{subsec:optgammandoptp}

\begin{figure}[t]
\centering
\includegraphics[width=0.45\textwidth]{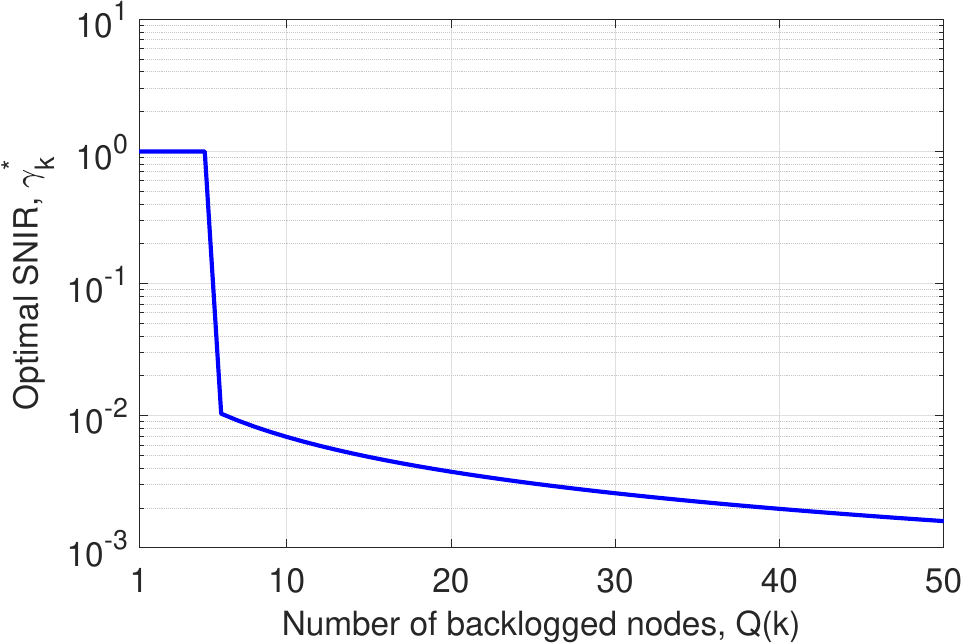}
\caption{Optimal target SNIR $\gamma^*_k$ (normalized with respect to $\gamma_{max}$), which maximizes the sum-rate, as a function of number of backlogged nodes $k$ ($\gamma_{\text{max}} = 31$, $\epsilon = 0.1$).}
\label{fig:gammastar_k}
\end{figure}

\begin{figure}[t]
\centering
\includegraphics[width=0.45\textwidth]{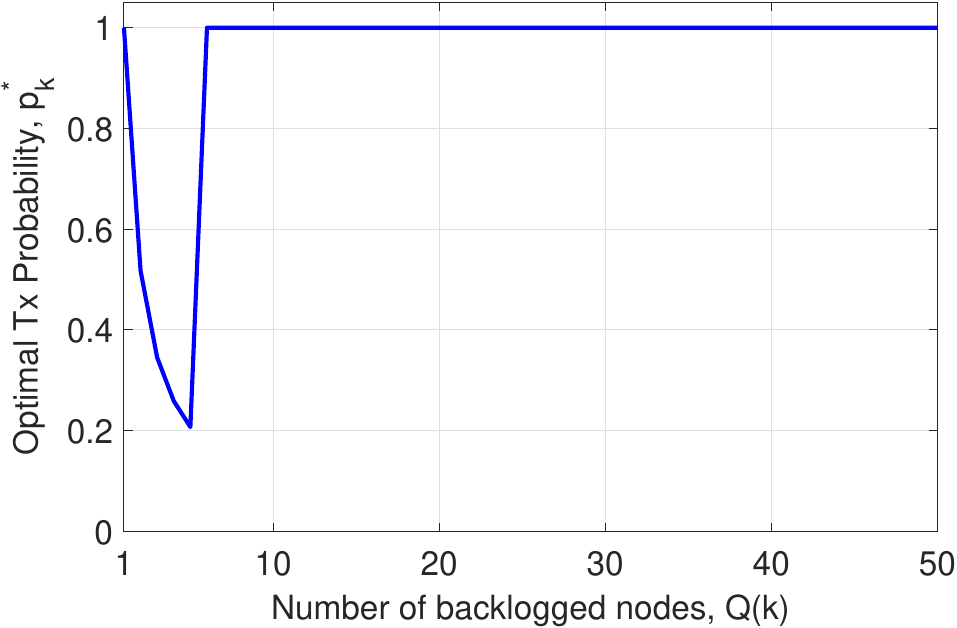}
\caption{Optimal transmission probability $p^*_k$, which maximizes the sum-rate, as a function of number of backlogged nodes $k$ ($\gamma_{\text{max}} = 31$, $\epsilon = 0.1$).}
\label{fig:pstar_k}
\end{figure}

\cref{fig:gammastar_k,fig:pstar_k} illustrates the optimal target \ac{SNIR} $\gamma^*_k$ and the optimal transmission probability $p^*_k$ respectively, as a function of the number of backlogged nodes $k$.
The values of $\gamma^*_k$ and $p^*_k$ are obtained by maximizing the system sum-rate $U_k(p,\gamma)$ (see \cref{eq:sumratexpression}), through numerical methods.
These values are obtained under the following conditions: (i) the maximum allowed \ac{SNIR} is equal to $\gamma_{\text{max}} = 31$; and (ii) the target success probability of detecting a packet is equal to $1-\epsilon = 0.9$, when there is only one node transmitting.
It is noted that $\gamma^*_k$ and $p^*_k$ plotted in \cref{fig:gammastar_k,fig:pstar_k} accurately match the simple expressions provided in \cref{eq:pstark,eq:gammastark}.

The optimal transmission probability decreases as the number of backlogged nodes grows, up to a breakpoint, $k_c$.
In the same range of values of $k$, the optimal \ac{SNIR} stays constant at its maximum allowed value $\gamma_{\text{max}}$.
This highlights that the MAC layer only allows a few nodes to transmit (the expected number of nodes transmitting in a slot is 1), yet with high spectral efficiency.

As the number of backlogged nodes increases beyond the critical point $k_c$, the optimal target \ac{SNIR} $\gamma^*_k$ starts decreasing, inversely proportional to $k$, while the optimal transmission probability is $1$.
In this scenario, the MAC layer allows every node to transmit, but with a low target \ac{SNIR}, resulting in low spectral efficiency, but large degree of parallelism in transmissions.

\begin{figure}[t]
\centering
\includegraphics[width=0.45\textwidth]{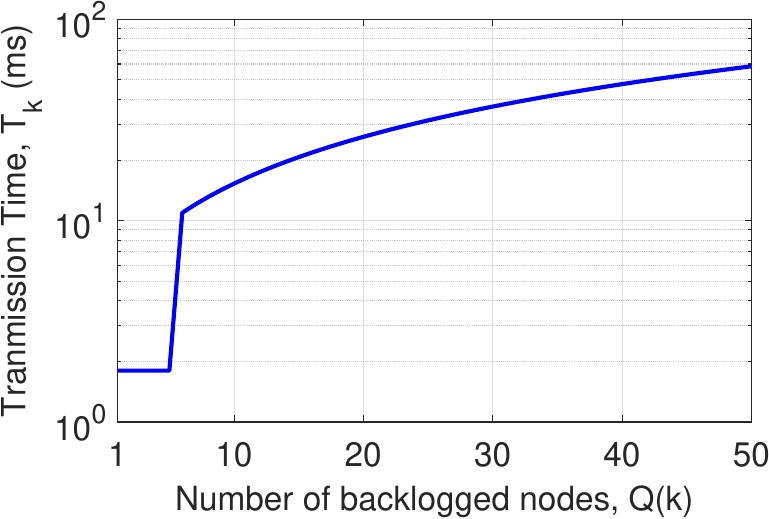}
\caption{Transmission time $T_k$ as a function of number of backlogged nodes $k$ ($\gamma_{\text{max}} = 31$, $\epsilon = 0.1$).}
\label{fig:transtime_k}
\end{figure}

The slot size $T_k$ is evaluated according to \cref{eq:Tkexpression}.
\cref{fig:transtime_k} shows the slot size $T_k$ as a function of the number of backlogged nodes.
For large $k$, since $\gamma^*_k \ll 1$, $T_k$ is inversely proportional to $\gamma^*_k$, hence proportional to $k$.
The slot size attains its minimum value, given that at least one node is backlogged, at $T_1 = T_{oh} + \frac{ L }{ W \log_2( 1 + \gamma_{\text{max}}) }$ for a number of backlogged nodes smaller than the critical threshold $k_c$.
With the numerical values in \cref{tab:1}, the minimum slot time, given that at least one node is backlogged, is $T_1 = 1.8$ ms.

\subsection{Access layer metrics}
\label{subsec:CBRandPDRandaccessdelay}

\begin{figure}[t]
\centering
\includegraphics[width=0.45\textwidth]{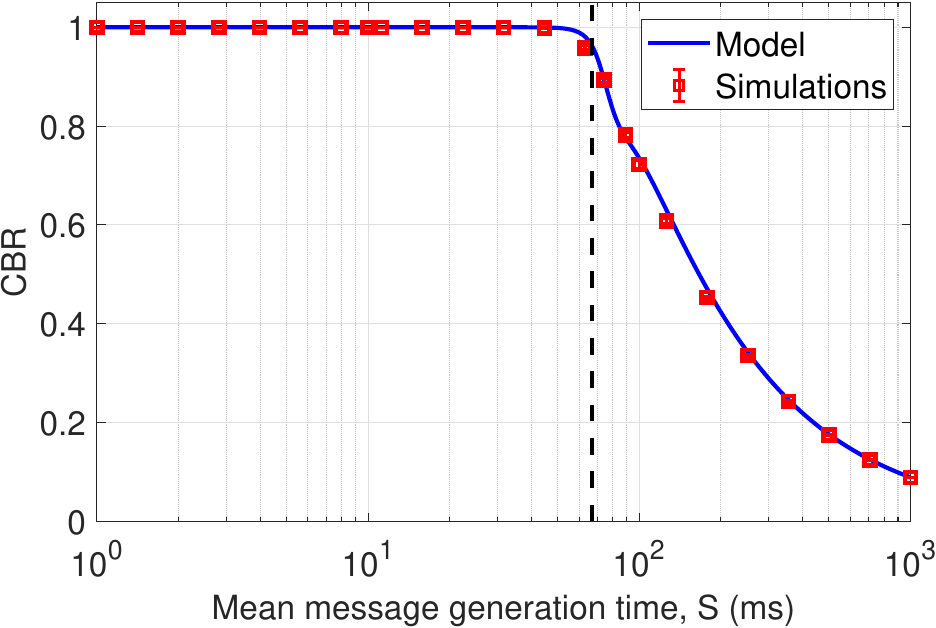}
\caption{Channel busy ratio as a function of mean message generation time, $S$.}
\label{fig:CBR_S}
\end{figure}

 Channel Busy Ratio (CBR) is shown in \cref{fig:CBR_S} as a function of mean message generation time $S$.
\ac{CBR} measures the mean fraction of time that the channel is busy due to transmission from any node.
As expected, \ac{CBR} is essentially $1$ in heavy traffic regime, while it drops quickly as the system moves to light traffic regime, where nodes are idle for most of their time.

\begin{figure}[t]
\centering
\includegraphics[width=0.45\textwidth]{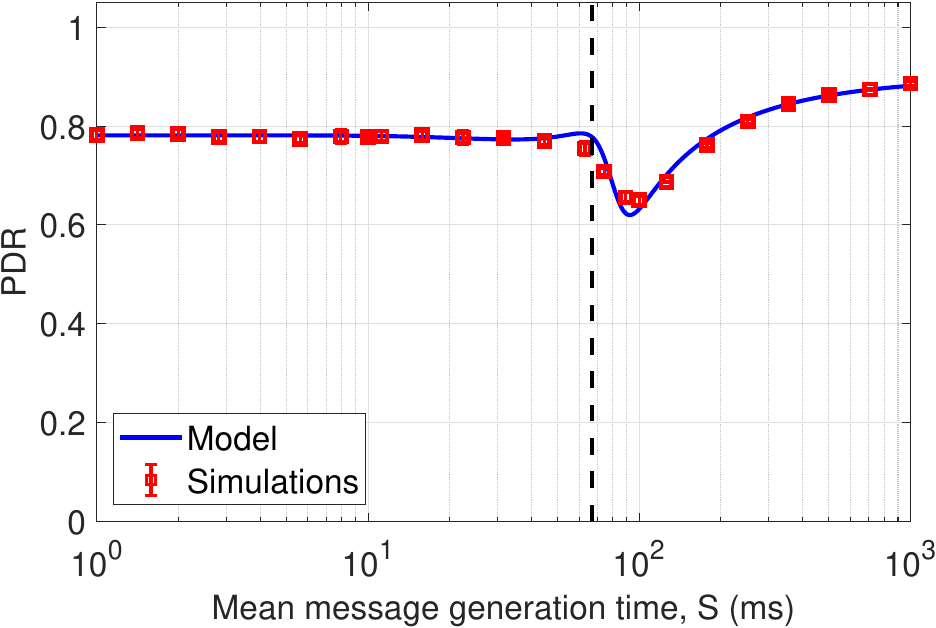}
\caption{Packet delivery ratio as a function of mean message generation time, $S$.}
\label{fig:PDR_S}
\end{figure}

Packet Delivery Ratio (PDR) is plotted in \cref{fig:PDR_S} as a function of $S$.
The \ac{PDR} measures the mean fraction of packets that are delivered successfully to the base station.
A high level of \ac{PDR} is achieved in the heavy traffic regime, which gives evidence of the effectiveness of \ac{SIC} in dealing with a large number of backlogged nodes.
%The cost involved includes both the time required for processing, as the slot time increases due to the need to decrease the target \ac{SNIR} to handle the increasing number of backlogged nodes, and the energy consumption by nodes involved in generating and transmitting new messages.

In the low traffic regime, the number of backlogged nodes is typically below $k_c$ and hence the transmission probability is set to $1/k$, when $k$ nodes are backlogged.
This entails that the expected number of transmitting nodes is $1$.
If more nodes transmit simultaneously, failure is most probable, given that $\gamma_k = \gamma_{\text{max}}$ in this region.
The system dynamic resembles that of classic Slotted ALOHA, wherein the success probability collapses as the load increases, i.e., $S$ decreases in our scenario, moving from large $S$ values on the right of the plot (light traffic regime) towards smaller values of $S$.
However, in the considered system, \ac{SIC} comes to rescue successful decoding of packets, thank to the adaptive access parameters.
As $S$ gets smaller and the load on the channel grows, the number of backlogged nodes gets larger than $k_c$ with higher and higher probability.
This triggers switching of $\gamma_k$ values from $\gamma_{\text{max}}$ to much smaller values (see \cref{eq:gammastark}).
Hence, success probability is restored and it settles to a quite large value ($\approx 0.89$) as we move towards the heavy traffic region.

The down notch seen in the \ac{PDR} plot is reminiscent of the performance drop of classic Slotted ALOHA.
When $S$ decreases from the right of the plot, the load on the system grows and the probability of failing decoding with $\gamma = \gamma_{\text{max}}$ grows.
While classic Slotted ALOHA throughput collapses as the load further increases, here the adaptation of the parameters $p$ and $\gamma$ restores high \ac{PDR} values, to the cost of slowing down transmission rate (longer time slots are used).

% While the packet delivery ratio slightly drops in the transition region, it then regains.
% It is important to note the meaningfulness of PDR in both regimes.
% In the heavy traffic regime, PDR is very high, but most nodes are backlogged and transmitting.
% Similarly, in the light traffic regime, PDR is high, but there are very few backlogged nodes.

\begin{figure}[t]
\centering
\includegraphics[width=0.45\textwidth]{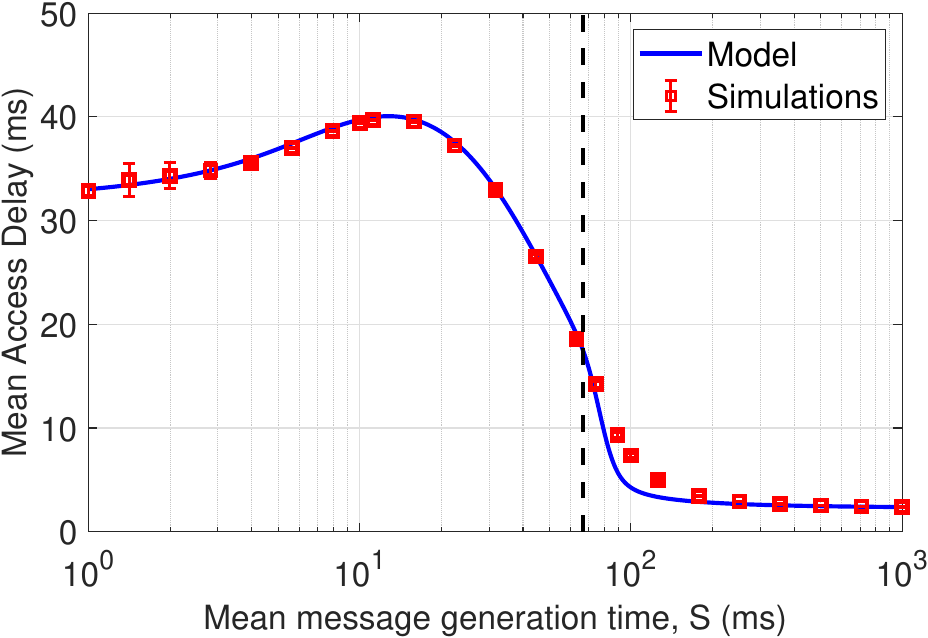}
\caption{Mean access delay as a function of mean message generation time, $S$.}
\label{fig:EAD_S}
\end{figure}

The mean access delay is shown in \cref{fig:EAD_S} as a function of mean message generation time $S$.
The mean access delay is defined as the time elapsing since when the message that will be transmitted arrives at the node, until when transmission of that message is complete.

In the heavy traffic regime, the mean access delay is very high, because the number of backlogged nodes $k$ is high, pushing the slot size $T_k$ to large values.
Smaller access delays are seen in the light traffic regime, mainly because of the much smaller slot size, even if using $p_k <1$ introduces a non-null contention time, i.e., on average $1/p_k$ time slots are required before transmission is carried out.

The central peak appearing in the curve of the access delay stems from the component $V$ of the access delay (see \cref{subsec:accessdelay}), i.e., the time elapsing since the last arrival of a new message and the end of the slot time where it occurs.
For small values of $S$ (hence, large values of $\lambda$), the last arrival in a slot occurs close to the end of the slot with high probability, so that $V$ is a small fraction of the slot time.
On the contrary, for large values of $S$ (hence, small values of $\lambda$), it is highly probable that $Q = 1$, hence the slot time is small, thus making $V$ again negligible.
In the central region, slot sizes are still quite large, but there are few arrivals in a slot most of the times, so that $V$ is in the same order of magnitude as the slot size and affects significantly the overall mean access delay.

\begin{figure}[t]
\centering
\includegraphics[width=0.45\textwidth]{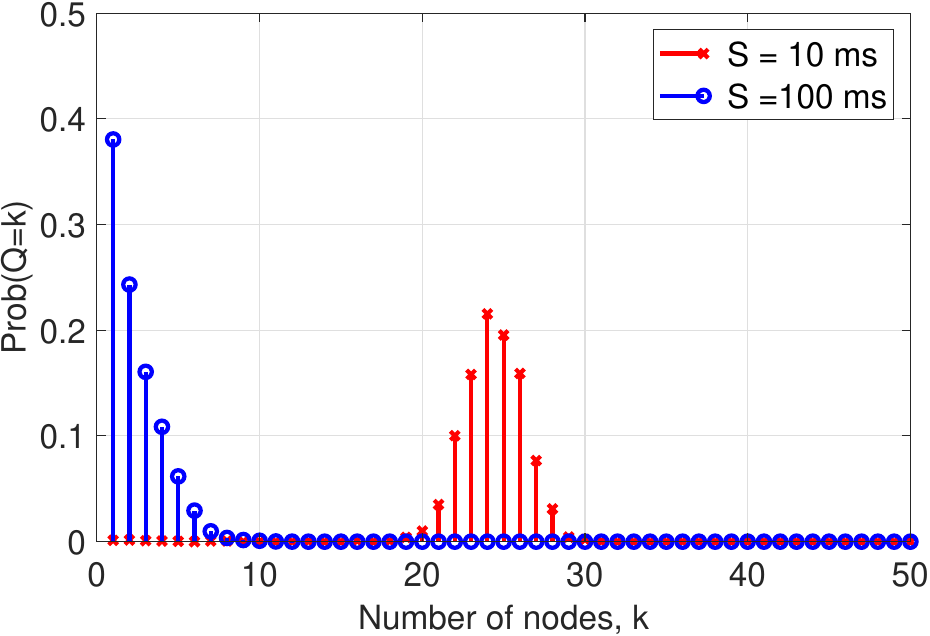}
\caption{Probability distribution of the number of backlogged nodes for two values of the mean message generation time $S$.}
\label{fig:Probability_Qk}
\end{figure}

More insight emerges from the analysis of the probability distribution of the number $Q$ of backlogged nodes, which is shown in \cref{fig:Probability_Qk} for two values of $S$.

For $S = 100 \text{ ms} > 1/\lambda_\infty$ (light traffic regime), the \ac{PDF} of $Q$ is concentrated over small values of $k$, i.e., few nodes are active in any slot with overwhelming probability.
On the opposite, for $S = 10 \text{ ms} < 1/\lambda_\infty$ (heavy traffic regime), the distribution of $Q$ tends to concentrate around the central part of the range $[0,n]$, resulting in a narrower peak as $S$ becomes smaller.
Inspection of simulation traces in heavy traffic reveals that nodes tend to split in two groups of comparable size, that transmit in alternating slot times.
For a sufficiently large number of nodes, both groups are large enough (with high probability) so that $\gamma_k$ is small and time slots are long.
Then, the system remains locked in this alternating behavior, where about half of the nodes transmit in one-time slot, while the other nodes wait to become backlogged and transmit in the next slot.

\begin{figure}[t]
\centering
\includegraphics[width=0.45\textwidth]{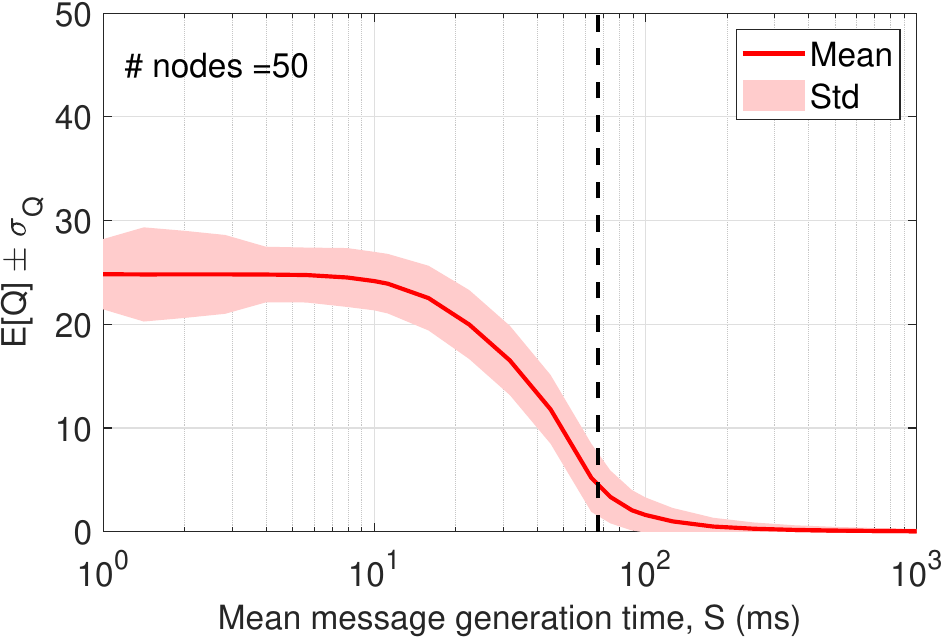}
\caption{Mean number of backlogged nodes, along with standard deviation, as a function of mean message generation time, $S$.}
\label{fig:mean_std_Qk}
\end{figure}

This alternating behavior of nodes can be seen from \cref{fig:mean_std_Qk}, where the mean number of backlogged nodes is shown as a function of mean message generation time $S$.
The shaded region accounts for a spread of one standard deviation of $Q$ across the mean.
The transition between heavy traffic to light traffic regime is apparent, as well as the fact that the sizes of the two node groups in heavy traffic are about the same ($\mathrm{E}[Q] \approx 25$ with $n = 50$ nodes).
Both the mean and the spread are much larger in heavy traffic, but this is handled nicely by \ac{SIC}, as far as data integrity is concerned.
The price to pay is an increase of delay, due to larger slot sizes.

\subsection{Application-related metrics}
\label{subsec:thruenergyAoI}

Throughput, mean consumed energy, and mean age of information are presented in this section. 

Throughput measures the mean delivered bit-rate or, if normalized, the mean fraction of generated messages that are successfully delivered to the base station.

Mean consumed energy gives the mean energy required per delivered packet.
Numerical values of the quantities used in this evaluation are listed in \cref{tab:1}.
The numerical values for energy consumption are loosely inspired to typical values for LoRaWAN equipment \cite{Casals2017}.

Age of information is the well known metric \cite{Yates2021}, referred to the age of the last current update data generated by each node and stored in the collecting base station.

\subsubsection{Throughput}

\begin{figure}[t]
\centering
\includegraphics[width=0.45\textwidth]{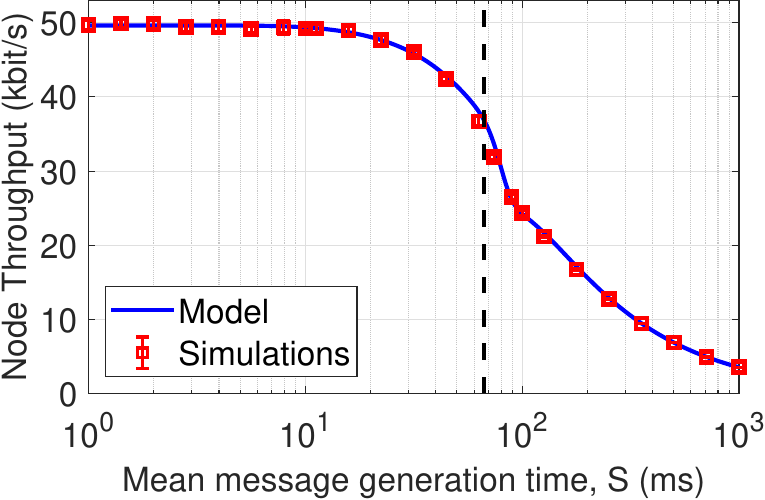}
\caption{Throughput in kilo-bit per second as a function of mean message generation time, $S$.}
\label{fig:thru_vs_S}
\end{figure}

\begin{figure}[t]
\centering
\includegraphics[width=0.45\textwidth]{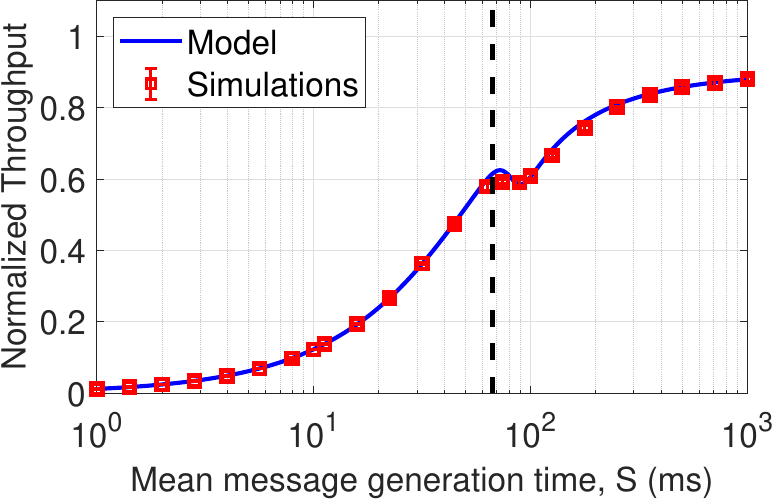}
\caption{Normalized throughput as a function of mean message generation time, $S$.}
\label{fig:nthru_vs_S}
\end{figure}

The node throughput and normalized throughput are shown in \cref{fig:thru_vs_S,fig:nthru_vs_S} respectively, as a function of mean message generation time $S$.
Node throughput in \cref{fig:thru_vs_S} is the mean carried bit rate of messages delivered to the base station, while in \cref{fig:nthru_vs_S} it is normalized with respect to message generation rate $\lambda = 1/S$.

The node throughput in \cref{fig:thru_vs_S} saturates when the system is pushed in the heavy traffic regime.
The adaptive multiple access scheme appears to scale robustly, with no collapse as the rate of generation of update messages increases in the limit for $S \rightarrow 0$.
On the opposite side, as $S$ grows, after the critical transition region, the node throughput falls, as expected, given the diminishing generation rate of new update messages.

\cref{fig:nthru_vs_S} shows that normalized throughput increases as the message generation time increases, dropping to negligible values for low values of $S$.
This behavior can be understood by analyzing the sources of message loss.
There are two sources of loss.
First, messages offered by the upper layer to the MAC entity are discarded, if the MAC entity is engaged in contention or in transmission.
Second, messages that are not decoded successfully, because of failure of \ac{SIC}, are lost as well.
The normalized throughput performance in the heavy traffic regime is primarily influenced by the first source of message loss.
Both sources of loss have comparable impact in the transition region between heavy and light traffic regimes.
Message loss is dominated by residual decoding errors when moving to the light traffic regime.
By the definition of power control, residual packet loss due to failure of decoding is $\epsilon = 0.1$.
Hence, the normalized throughput tends to 0.9 as $S$ tends to infinity ($\lambda$ tends to 0), so that each node transmits alone with high probability.

Summing up, the two throughput plots suggest that the highest possible throughput rate (saturation) is achieved under heavy traffic, which however entails that most of the generated messages are discarded before any transmission attempt occurs.
The highest efficiency, indicated by a large \emph{normalized} throughput, is achieved under the light traffic regime.
However, the throughput rate is relatively low in this regime
A compromise is struck around the critical point of transition between the two regimes, where a high throughput and a relatively high efficiency of delivery of generated messages can be achieved.

\subsubsection{Age of Information}

\begin{figure}[t]
\centering
\includegraphics[width=0.45\textwidth]{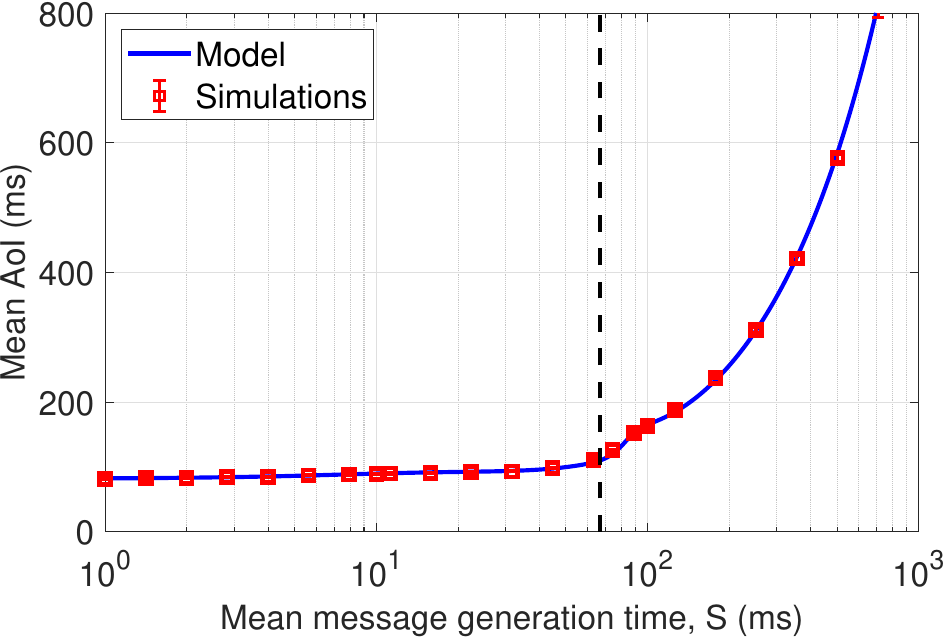}
\caption{Mean AoI as a function of mean message generation time, $S$}
\label{fig:aoi_vs_S}
\end{figure}

The mean \ac{AoI} is shown in \cref{fig:aoi_vs_S} as a function of mean message generation time $S$.
In the light-traffic regime, the mean \ac{AoI} grow steeply as $S$ gets larger, thus making new update message generation more and more slack.
As we move to a heavy-traffic regime, the \ac{AoI} attains its minimum, since update messages are generated more frequently and \ac{SIC} helps relieving the congestion on the channel.
The optimal operation region in terms of the mean \ac{AoI} is the heavy-traffic regime.
This is consistent with throughput (in terms of bit/sec).
However, in the heavy-traffic regime we have seen that normalized throughput is low and access delay is large.
We will see in next subsection that an energy price is paid as well, thus giving rise to a trade-off between \ac{AoI} and mean consumed energy.

\subsubsection{Energy}

\begin{figure}[t]
\centering
\includegraphics[width=0.45\textwidth]{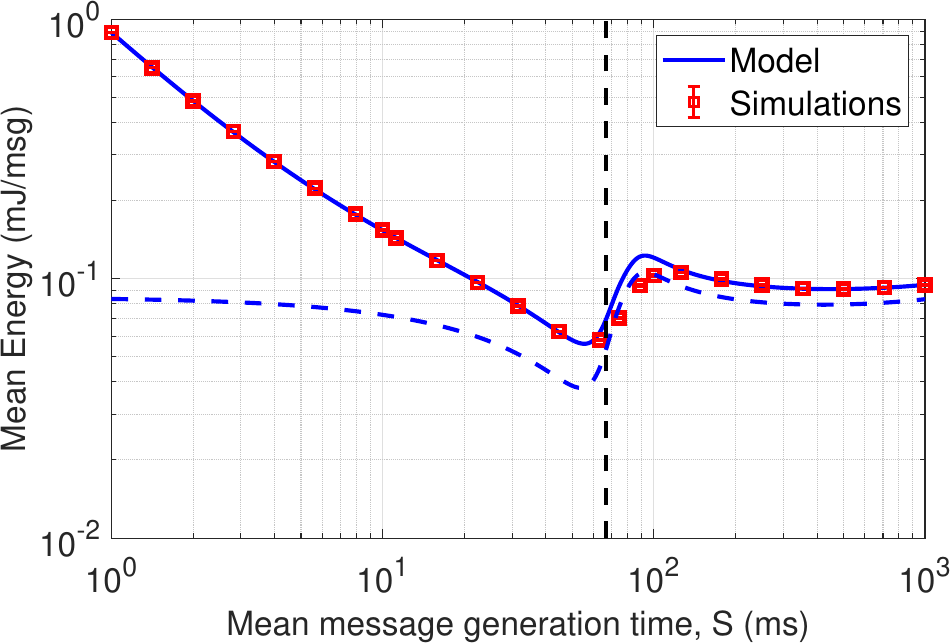}
\caption{Mean energy as a function of mean message generation time, $S$.}
\label{fig:mean_energy_vs_S}
\end{figure}

The mean energy spent per delivered message is shown in \cref{fig:mean_energy_vs_S} as a function of mean message generation time $S$.
This metric accounts for energy spent for generating new messages, for powering up the node, and for message transmission.
The blue dotted line represents the mean energy cost per delivered packet, without accounting for energy consumed in new message generation.

In the light-traffic regime, the mean energy is relatively low, as nodes doze for most of the time, while message transmissions are successful with high probability.

As we move towards the heavy traffic regime, to the left of the plot (small values of $S$), the energy spent per delivered message is dominated by the cost of generating new messages, most of which are actually discarded at the node, without being transmitted.
If we neglected this energy cost, the mean amount of energy spent per delivered packet would saturate in heavy traffic (see the dotted line in \cref{fig:mean_energy_vs_S}), highlighting the effectiveness of \ac{SIC}.
This conclusion is overturned by the MAC layer policy to discard new message arriving when there is already one message being dealt with.
On the other hand, storing all newly generated messages for later transmission is not beneficial to \ac{AoI}. 
As a result, several generated messages must be discarded, which puts a penalty on energy consumption, with no benefit to \ac{AoI}.

\subsection{Trade-off  between AoI and Energy}

\begin{figure}[t]
\centering
\includegraphics[width=0.45\textwidth]{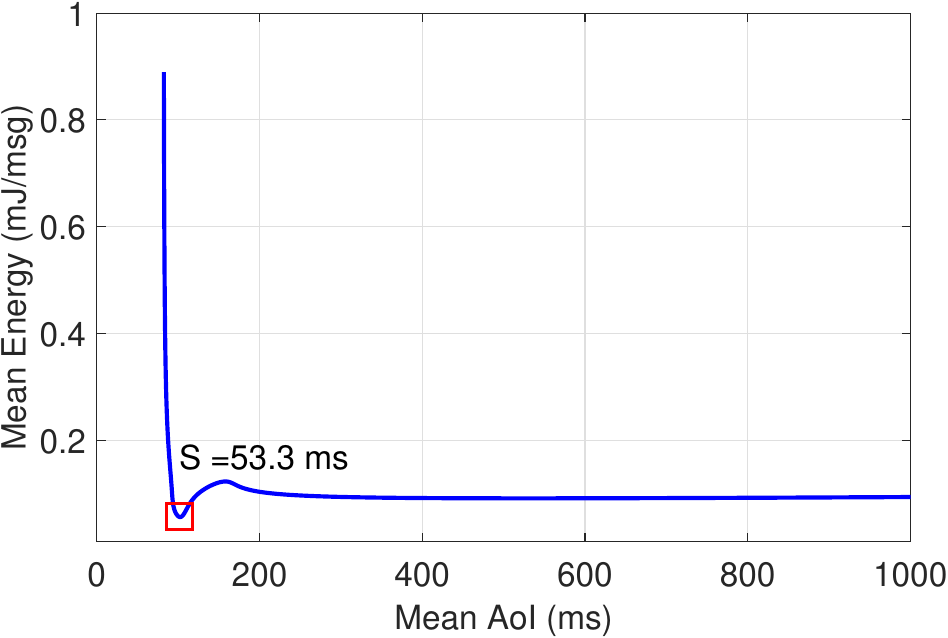}
\caption{Mean energy vs mean \ac{AoI} (both as a function of mean message generation time S).}
\label{fig:energy_vs_aoi}
\end{figure}

The trade-off of mean energy required per delivered message against the mean \ac{AoI} is plotted in \cref{fig:energy_vs_aoi}.
The trade-off is obtained by varying the mean message generation time, $S$.
The obtained trade-off exhibits a very sharp knee of the curves that appears in the low-left corner of the graph, identifying the optimal operating point in terms of energy and \ac{AoI} highlighted in \cref{fig:energy_vs_aoi}.
With the given numerical values of parameters, the optimal working point is achieved when the $50$ nodes generate new update messages with an average period of about $53$ ms.
The corresponding values of mean \ac{AoI} and mean required energy per delivered message are $101$ ms and $0.06$ mJ/msg, respectively.
If we try to reduce the mean \ac{AoI}, we face a very steep increase in the energy cost for minimal improvements in data freshness.
Conversely, any further reduction of mean required energy impacts heavily on \ac{AoI}, in spite of minimal obtained energy gains.

% -------------- Section end marker --------------
%                _       _
%               ( )_    ( )
%    ___  _   _ | ,_)   | |__     __   _ __   __
%  /'___)( ) ( )| |     |  _ `\ /'__`\( '__)/'__`\
% ( (___ | (_) || |_    | | | |(  ___/| |  (  ___/
% `\____)`\___/'`\__)   (_) (_)`\____)(_)  `\____)
%
% -------------- Section end marker --------------

\section{Conclusions}
\label{sec:conclusions}

We have addressed a model for a multiple access system designed for nodes that share a common communication channel to send update messages to a central base station.
The motivation of the model is to understand the interplay between \ac{SIC} and multiple access, exploring the performance achievable under an ideal \ac{SIC} receiver, by adapting the multiple access parameter to the number of backlogged nodes.

An analytical model of the system has been defined, based on mean field approximation.
The proposed analytical model turns out to be highly accurate and yet simple.
It allows computing all relevant metrics, including the \ac{PDF} of main system performance metrics.
Moreover, the proposed model is conducive to several generalizations.
Possible lines of further research can be identified as follows.

Statistical estimators of the number of backlogged nodes should be designed to drive the parameters $p$ and $\gamma$ as a function of the (estimated) number of backlogged nodes.
More generally, a reinforcement learning approach could be adopted to set $p$ and $\gamma$, especially in a time-varying operational environment where the number of nodes or the traffic patterns of nodes vary with time.

The model can be adapted to account for packet re-transmissions, assuming feedback is provided by the base station.
It can also be generalized to heterogeneous nodes with different generation rates, e.g., two classes of nodes with different values of $\lambda$. 
Correspondingly, we would have two different values of the probability of being backlogged and hence a system of two nonlinear fixed point equations.

On the theoretical side, it is interesting to characterize the asymptotic behavior of the model and hence the achievable performance as the number of nodes grows.
The core point here is to characterize \ac{SIC} outcome as $n$ scales.

From an implementation standpoint, it is crucial to define and experiment with practical coding techniques and signal processing algorithms that support \ac{SIC}, thus identifying how much of the promised performance can be reaped in spite of the non-idealities of practical \ac{SIC} implementation.

Finally, from an application point of view, the proposed model could be the heart of a larger model to investigate how distributed control applies through the interaction of sensors collecting measurements and reporting them to a controller that runs actuators.
This is a classic closed-loop control with feedback delay, where the studied multiple access system is inserted in the feedback loop.

% -------------- Section end marker --------------
%                _       _
%               ( )_    ( )
%    ___  _   _ | ,_)   | |__     __   _ __   __
%  /'___)( ) ( )| |     |  _ `\ /'__`\( '__)/'__`\
% ( (___ | (_) || |_    | | | |(  ___/| |  (  ___/
% `\____)`\___/'`\__)   (_) (_)`\____)(_)  `\____)
%
% -------------- Section end marker --------------
\appendices

\section{Probability distributions and moments of $C$ and $R$}
\label{app:A}

Expressions of the Laplace transforms of the \acp{PDF} and of the moments of $C$ and $R$ are given in this Appendix.

As for $C$, let us condition on the event $Q = k$, i.e., there are $k$ nodes backlogged, besides the tagged one.
Then, we have:
\begin{equation}
%\label{ }
C |_{Q=k} = \begin{cases}
     X^\prime |_{Q=k} = T_{k+1} & \text{w.p. } p_{k+1}, \\
     X^\prime |_{Q=k}+\tilde{C} = T_{k+1} + \tilde{C} & \text{otherwise}.
\end{cases}
\end{equation}
where $\tilde{C}$ is the \emph{residual} time required for the tagged node to make its transmission.
Since the process is memory-less, because of the mean field approximation, the random variable $\tilde{C}$ has the same \ac{PDF} as $C$.
Hence:
\begin{align}
%\label{ }
\varphi_{C |_{Q=k}}(s) &= \mathrm{E}[ e^{ - s C } | Q = k ]   \nonumber \\
  &= p_{k+1} e^{ - s T_{k+1} } + (1-p_{k+1} ) e^{ - s T_{k+1} } \varphi_C(s)
\end{align}
for $k = 0,\dots,n-1$.
Removing the conditioning and solving for $\varphi_C(s)$, we get:
\begin{equation}
\varphi_{C}(s) = \frac{ \sum_{ k = 0 }^{ n-1 }{ q_k p_{k+1} e^{ - s T_{k+1} } } }{ 1 - \sum_{ k = 0 }^{ n-1 }{ q_k (1-p_{k+1}) e^{ - s T_{k+1} } } }
\end{equation}
The mean value of $C$ is:
\begin{equation}
%\label{ }
\mathrm{E}[ C ] = \frac{ \sum_{ k = 0 }^{ n-1 }{ q_k T_{k+1} } }{ \sum_{ k = 0 }^{ n-1 }{ q_k p_{k+1} } } = \frac{ \overline{T}^\prime }{ \overline{p}^\prime }
\end{equation}
where $ \overline{T}^\prime$ and $ \overline{p}^\prime$ are the mean slot time and the probability of transmission of a slot where the tagged node is backlogged.
The second moment of $C$ is found by deriving twice $\varphi_C(s)$:
\begin{equation}
%\label{ }
\mathrm{E}[ C^2 ] = \frac{ \sum_{ k = 0 }^{ n-1 }{ q_k T_{k+1}^2 } + 2 \mathrm{E}[ C ] \sum_{ k = 0 }^{ n-1 }{ q_k ( 1 - p_{k+1} ) T_{k+1} } }{ \sum_{ k = 0 }^{ n-1 }{ q_k p_{k+1} } }
\end{equation}

The time $R$ elapsed from the departure of a packet until the beginning of the contention time of the next arriving packet is given by:
\begin{equation}
\label{eq:Rdefinition}
R = \sum_{ i=1 }^{ N }{ X(i) }
\end{equation}
where $N$ is a discrete random variable, defined as the number of slot times until a new packet arrives at the tagged node.

% HERE IS AN ALTERNATIVE DERIVATION OF $\varphi_R(s)$.
%\begin{equation}
%\label{ }
%\left. R \right|_{Q = k} = \begin{cases}
%    T_k  & \text{w.p. } 1-e^{ - \lambda T_k }, \\
%    T_k + \tilde{R}  & \text{w.p. } e^{ - \lambda T_k }
%\end{cases}
%\end{equation}

From the definition in \cref{eq:Rdefinition}, conditional on $N = h$ and on $X(j) = x_i, \, i =1,\dots,h$, it is $R = x_1+\dots,x_h$.
Hence:
\begin{equation*}
\mathrm{E}[ e^{-sR} | N = h, X(1) = x_1,\dots,X(h) = x_h ] = \prod_{ j = 1 }^{ h }{ e^{ - s x_j } }.
\end{equation*}
Since arrivals occur according to a Poisson process of mean rate $\lambda$, the probability of $N = h$, conditional on $X(j) = x_j, \, j =1,\dots,h$, is $e^{-\lambda x_1} \dots e^{-\lambda x_{h-1} } (1-e^{-\lambda x_h})$.
Removing the conditioning we have:
\begin{align}
\varphi_R(s) &= 
    \sum_{ h = 1 }^{ \infty} \int_{ 0 }^{ \infty } (1-e^{-\lambda x_h}) f_X(x_h) e^{ - s x_h } dx_h  \nonumber\\
    & \qquad\times \prod_{ j = 1 }^{ h-1 }{ \int_{ 0 }^{ \infty }{ e^{ - \lambda x_j } f_X(x_j) e^{ - s x_j } } dx_j } \nonumber\\
    &= \sum_{ h = 1 }^{ \infty}{ [\varphi_X(s+\lambda)]^{h-1} \left[ \varphi_X(s) - \varphi_X(s+\lambda) \right] } \nonumber\\
    &= \frac{ \varphi_X(s) - \varphi_X(s+\lambda) }{ 1 - \varphi_X(s+\lambda) }.
\end{align}
where $\varphi_X(s)$ is given in \cref{eq:LTPDFXdefine}.

The first two moments of $R$ are as follows:
\begin{align}
%\label{}
    &\mathrm{E}[ R ] = \frac{ \mathrm{E}[ X ] }{ 1- \varphi_X( \lambda ) }   \\
    &\mathrm{E}[ R^2 ] = \frac{ \mathrm{E}[ X^2 ] }{ 1- \varphi_X( \lambda ) }  - \frac{ 2 \mathrm{E}[ X ] \varphi_X^\prime( \lambda ) }{ ( 1- \varphi_X( \lambda ) )^2 } 
\end{align}
where $\varphi_X^\prime(\cdot)$ denotes the derivative of $\varphi_X(\cdot)$.

\section{Probability distributions of $N$ and $M$}
\label{app:B}

Let $N$ be the number of slots it takes for a new message to be generated at the tagged node, after the tagged node has transmitted the previous message.
The probability of no arrival in a slot time at the start of which the tagged node is idle is given by $\varphi_X(\lambda)$.
Then, it is easy to check that $N$ has the following Geometric probability distribution:
\begin{equation}
\label{eq:NPMFdefine}
\mathcal{P}( N = h ) = \left[ 1 - \varphi_X(\lambda)  \right] \left[ \varphi_X(\lambda) \right]^{h-1} \, , \; h \ge 1.
\end{equation}

Let $M$ be the number of slots it takes for the tagged node to complete its transmission (including contention).
Conditioning on $Q = k$, we get:
\begin{equation}
%\label{ }
M |_{Q=k} = \begin{cases}
     1 & \text{w.p. } p_{k+1}, \\
     1 + \tilde{M} & \text{otherwise}.
\end{cases}
\end{equation}
where $\tilde{M}$ is the residual node busy time.
Thanks to the mean field approximation, $\tilde{M}$ has the same probability distribution as $M$.
Hence:
\begin{align}
%\label{ }
\phi_{M |_{Q=k}}(z) &= \mathrm{E}[ z^{M} | Q = k ]   \nonumber  \\
  &= p_{k+1} z + (1-p_{k+1}) z \phi_M(z)
\end{align}
Removing the conditioning and solving for $\phi_M(z)$, we get:
\begin{equation}
%\label{ }
\phi_{M}(z) = \frac{ \sum_{ k = 0 }^{ n-1 }{ q_k p_{k+1} } \, z }{ 1 - \sum_{ k = 0 }^{ n-1 }{ q_k (1-p_{k+1}) } z } = \frac{ \overline{p}^\prime z }{ 1 - ( 1 - \overline{p}^\prime ) z }
\end{equation}
where $\overline{p}^\prime = \sum_{ k = 0 }^{ n-1 }{ q_k p_{k+1} }$ is the mean probability of transmission in a slot where the tagged node is backlogged.
Inverting the generating function of $M$, it is found that:
\begin{equation}
%\label{ }
\mathcal{P}( M = m ) = \overline{p}^\prime \left( 1 - \overline{p}^\prime \right)^{m-1} \, , \quad m \ge 1
\end{equation}

\section{Proof of uniqueness of fixed point of \cref{eq:bfixedpoint}}
\label{app:C}

The fixed point \cref{eq:bfixedpoint} can be written as:
\begin{equation}
\label{eq:fixedpointappendix}
b = \frac{ 1 - \varphi_X(\lambda) }{ 1 + \overline{p}^\prime - \varphi_X(\lambda) } = \frac{ \sum_{ k = 0 }^{ n-1 }{ a_k q_k } }{ \sum_{ k = 0 }^{ n-1 }{ (a_k + p_{k+1} ) q_k } } = F(b)
\end{equation}
where $a_k = 1-e^{ - \lambda T_k }, \; k = 0,1,\dots,n-1$. 
The function $F(b)$ is analytic in $(0,1)$, with $0 < F(0) = a_0/(a_0+1) < 1$ and $0 < F(1) = a_{n-1}/(a_{n-1}+p_n) < 1$.
Brouwer's theorem guarantees that a fixed point exists and it must belong to the interior of the interval $[0,1]$ given that $F(b)$ is strictly positive and less than 1 at the extremes of the interval.

Let $c_k = \binom{n-1}{k}$ and $u = b/(1-b)$, whence $b = u/(1+u)$.
Then \cref{eq:fixedpointappendix} can be re-written as follows:
\begin{equation}
\frac{ u }{ 1+u } = \frac{ \sum_{ k = 0 }^{ n-1 }{ a_k c_k u^k } }{ \sum_{ k = 0 }^{ n-1 }{ (a_k + p_{k+1} ) c_k u^k } }
\end{equation}
Re-arranging terms, we get:
\begin{equation}
\label{eq:polnomialequalitywithu}
\sum_{ k = 0 }^{ n-1 }{ p_{k+1} c_k u^{k+1} } = \sum_{ k = 0 }^{ n-1 }{ a_k c_k u^k }
\end{equation}

Since $b \in [0,1]$, $u$ is restrained to the positive semi-axis.

The probabilities $p_k$ are set according to \cref{eq:pstark} for $k = 1,\dots,n$, that is to say, $p_k = 1/k$ for $k = 1,\dots, k_c-1$ and $p_k = 1$ otherwise.
The parameter $k_c$ is a given positive integer.

Let us first tackle the special case where $n \le k_c-1$.
Since $\gamma_k = \gamma_{\text{max}}$ for $k = 1,\dots,k_c - 1$, in case $n \le k_c-1$, we have $a_k = a_1$ for all $k = 1,\dots,n-1$.
Exploiting this special values of the $a_k$'s, together with $p_{k+1} = 1/(k+1)$ for $k = 0,\dots,n-1$, the left hand side of \cref{eq:polnomialequalitywithu}, denoted with $A(u)$, is cast into the following special form for $n \le k_c - 1$:
\begin{equation}
A(u) = \sum_{ k = 0 }^{ n-1 }{ \frac{ 1 }{ k+1 } c_k u^{k+1} } = \frac{ 1 }{ n } \left[ ( 1+u )^n - 1 \right]
\end{equation}
As for the right-hand side of \cref{eq:polnomialequalitywithu}, denoted with $B(u)$, we get:
\begin{equation}
B(u) = a_0 + a_ 1 \sum_{ k = 1 }^{ n-1 }{ c_k u^k } = a_0 + a_1 ( 1+u )^{ n-1 } - a_1
\end{equation}

It is $A(0) = 0 < a_0 = B(0)$.
Moreover, it is $A(u) > B(u)$ for sufficiently large values of $u$ since $A(u)$ is a polynomial of degree $n$, while $B(u)$ is a polynomial of degree $n-1$.
% \footnote{As a matter of example, for $n > 1$, it is $A(u) > B(u)$ for $u = n(a_0+1)-1$.}
Since both $A(u)$ and $B(u)$ are strictly increasing and convex for $u \ge 0$, it follows that there exists a unique intersection of the curves of $A(u)$ and $B(u)$ for some positive $u^*$.
This completes the proof in case $n \le k_c - 1$. 

In the following, we assume therefore $n \ge k_c$.
Then, \cref{eq:polnomialequalitywithu} can be re-written as follows:
\begin{equation}
\sum_{ k = 0 }^{ n-1 }{ c_k u^{k+1} } = \sum_{ k = 0 }^{ k_c-1 }{\frac{ k }{ k+1 } c_k u^{ k+1 } } + \sum_{ k = 0 }^{ n-1 }{ a_k c_k u^k }
\end{equation}

The left-hand side can be expressed in closed form, using the binomial expansion formula, obtaining $\sum_{ k = 0 }^{ n-1 }{ c_k u^{k+1} }  = u ( 1 + u )^{n-1} = A(u)$.

The right-hand side is a polynomial $B(u)$ of degree $n-1$ with positive coefficients.

It is easily verified that both $A(u)$ and $B(u)$ are strictly increasing and strictly convex for $u \ge 0$.
Moreover, it is $A(0) = 0 < a_0 = B(0)$ and $A(u) > B(u)$ for sufficiently large $u$, since $A(u) \sim u^n$ while $B(u) \sim u^{n-1}$, as $u \rightarrow \infty$.
It follows that there is a unique value of $u$, say $u^* > 0$, such that $A(u) = B(u)$.
The corresponding value of $b$, i.e., $b^* = u^*/(1+u^*)$ is the unique solution of \cref{eq:fixedpointappendix} in $(0,1)$.

\section*{Acknowledgement}

This work was partially supported by the European Union under the Italian National Recovery and Resilience Plan (NRRP) of Next Generation EU, partnership on ``Telecommunications of the Future'' (PE00000001 - program ``RESTART'').

\newpage

%\vspace{-33pt}
%\begin{IEEEbiography}[{\includegraphics[width=1in,height=1.25in,clip,keepaspectratio]{fig1}}]{Andrea Baiocchi}
%Use $\backslash${\tt{begin\{IEEEbiography\}}} and then for the 1st argument use $\backslash${\tt{includegraphics}} to declare and link the author photo.
%Use the author name as the 3rd argument followed by the biography text.
%\end{IEEEbiography}

%\vspace{-33pt}
%\begin{IEEEbiography}[{\includegraphics[width=1in,height=1.25in,clip,keepaspectratio]{fig1}}]{Asmad Razzaque}
%Use $\backslash${\tt{begin\{IEEEbiography\}}} and then for the 1st argument use $\backslash${\tt{includegraphics}} to declare and link the author photo.
%Use the author name as the 3rd argument followed by the biography text.
%\end{IEEEbiography}
%\vspace{11pt}
\begin{IEEEbiographynophoto}{Asmad Bin Abdul Razzaque}
(Graduate Student Member, IEEE) received the B.Sc. degree in Electrical Engineering from the University of Engineering and Technology (UET) and the M.Sc. degree in Electrical Engineering with a major in Telecommunications from the National University of Sciences and Technology (NUST), Pakistan. Currently, he is pursuing a Ph.D. in Information and Communications Technology (ICT) at the University of Rome Sapienza.
To date, he has published a few papers and has also received a Best Paper Award in an IEEE conference for his contributions.
His research is focused on advanced techniques in signal processing for telecommunications.
\end{IEEEbiographynophoto}

\vskip -2\baselineskip plus -1fil

\begin{IEEEbiographynophoto}{Andrea Baiocchi}
(Member, IEEE) received the Laurea degree in electronics engineering and the Ph.D. degree in information and communications engineering from the University of Rome La Sapienza, in $1987$ and $1992$, respectively.
Since January $2005$,he has been a Full Professor at the Department of Information Engineering, Electronics and Telecommunications, University of Rome Sapienza.
His main scientific contributions are on telecommunications network traffic engineering, queuing theory, resource sharing, and congestion control.
His research activities have been carried out also in the framework of many national (CNR, MIUR, and POR) and international (European Union and ESA) projects, also taking coordination and responsibility roles. He has published more than $170$ papers on international journals and conference proceedings.
He is the author of the book Network Traffic Engineering–Stochastic Models and Applications (Wiley, $2020$). He has participated to the Technical Program Committees of $80$ international conferences.
His current research interests focus on massive multiple access and vehicular networking. He also served in the Editorial Board of the Telecommunications technical journal [Telecom Italia (currently TIM)] for ten years.
He is currently an Associate Editor of Elsevier Vehicular Communications journal.
\end{IEEEbiographynophoto}

\end{document}